\documentclass[journal,twoside,web]{ieeecolor}
\usepackage{generic}
\usepackage{cite}
\usepackage{amsmath,amssymb,amsfonts}
\usepackage{graphicx}
\usepackage{textcomp}
\usepackage{comment}
\usepackage{xcolor}
\usepackage{verbatim}

\usepackage{mathrsfs}
\usepackage{placeins}
\usepackage[ruled,vlined]{algorithm2e}

\usepackage{subfig}
\usepackage{circuitikz}
\usetikzlibrary{fit}  
\usepackage{nicematrix,booktabs}
\def\BibTeX{{\rm B\kern-.05em{\sc i\kern-.025em b}\kern-.08em
    T\kern-.1667em\lower.7ex\hbox{E}\kern-.125emX}}
\begin{document}
\title{DTMC-Based Analysis and Scheduling for Periodic Flows with Proactive HARQ}
\author{%
\mbox{Haozhe Yi\textsuperscript{1,2}},\,
\mbox{Junyi Liu\textsuperscript{3,*}},\,
\mbox{Maolin Yang\textsuperscript{1}},\,
\mbox{Haochun Liang\textsuperscript{1,3}},\
\mbox{Bo Liu\textsuperscript{4}},\
\mbox{Feng Hong\textsuperscript{5}}%
\\[-1pt]
\makebox[\textwidth][c]{%
\mbox{Chaowei Liu\textsuperscript{4}},\
\mbox{Hongbiao Liu\textsuperscript{4}}%
}%
\thanks{\textsuperscript{1}University of Electronic Science and Technology of China, Chengdu, China.}%
\thanks{\textsuperscript{2}National University of Singapore, Singapore.}%
\thanks{\textsuperscript{3}Northeastern University, Shenyang, China.}%
\thanks{\textsuperscript{4}Beijing Institute of Control Engineering, Beijing, China.}%
\thanks{\textsuperscript{5}Beijing Institute of Computer Technology and Applications, Beijing, China.}%
\thanks{\textsuperscript{*}Corresponding author.}%
}

\maketitle

\begin{abstract}
Ultra-Reliable Low-Latency Communication (URLLC) requires strict reliability and latency guarantees for heterogeneous periodic traffic. Proactive HARQ improves resource efficiency through early termination, but slot-level timing effects, particularly delayed feedback, complicate schedulability analysis.

This paper presents a discrete-time Markov chain (DTMC)-based framework for periodic flows with proactive HARQ. By expanding the state space, the model captures HARQ round-trip time and other cross-slot timing effects. The framework determines the transmission opportunities required to satisfy heterogeneous reliability and latency constraints and supports offset-based scheduling through a two-stage genetic algorithm.

Simulations with industrial URLLC traffic show that the proposed method achieves higher schedulability than reactive HARQ, K-Repetition, and non-guaranteed proactive HARQ, with acceptable computational overhead.
\end{abstract}


\section{Introduction}
\label{sec:introduction}

Ultra-Reliable Low-Latency Communication (URLLC) is a key service category of 5G New Radio (NR), with stringent latency and reliability requirements. A representative target defined by 3GPP is a packet error rate of \(10^{-5}\) for a 32-byte payload within 1~ms user-plane latency, while future industrial applications may require reliability up to \(10^{-6}\) and latency below 1~ms \cite{3gpp38300,3gpp38913}. Such requirements are particularly important in cyber-physical systems (CPS), where industrial applications such as cooperative robotics, mobile operation panels, and emergency-stop services generate heterogeneous periodic traffic with different reliability and deadline requirements while competing for limited communication resources. Since these communications directly affect sensing, control, and actuation \cite{3gpp22104}, failure to satisfy the requirement of any flow may degrade system performance or compromise safety.

Achieving these stringent requirements is not straightforward and relies on several supporting technologies. To reduce transmission latency, recent 5G releases have introduced semi-persistent scheduling (SPS), which preconfigures recurring downlink transmission resources through one-time radio resource control signaling. On the one hand, SPS avoids the per-packet scheduling requests and access delays associated with dynamic scheduling; on the other hand, its periodic resource configuration is well suited to periodic URLLC traffic.

For reliability enhancement, retransmission is an important mechanism. However, in preconfigured resource allocation, too few retransmission opportunities can result in insufficient reliability, while excessive reservation leads to systematic resource waste. In Release 16, proactive HARQ (P-HARQ) addresses this tradeoff by configuring multiple transmission opportunities in advance while terminating the remaining transmissions once successful decoding becomes observable \cite{3gpp_ran1_2017,liu2020analyzing}. This early-termination mechanism preserves low latency and enables unused opportunities to be reused by subsequent packets. Nevertheless, the actual release time cannot be determined solely by the decoding time. In 5G NR, packet transmission, receiver processing, feedback generation, and feedback reception may span multiple slots. Consequently, the transmitter may continue sending redundant transmissions before the ACK becomes observable, complicating reliability analysis and resource configuration.

This paper studies how to configure proactive-HARQ transmission opportunities for heterogeneous periodic flows sharing the same resource. We develop a discrete-time Markov chain (DTMC)-based framework that captures HARQ round-trip time and other cross-slot timing effects through state-space expansion. The model determines the transmission opportunities required for each packet to satisfy its reliability and latency constraints. Based on this analysis, we propose a two-stage scheduling algorithm that first tests the synchronous-release configuration and then applies genetic-algorithm-based offset search when necessary.

We evaluate the proposed method using industrial URLLC traffic settings derived from 3GPP use cases \cite{3gpp22104}. The results show that it improves schedulability over existing retransmission and configuration approaches while maintaining acceptable computational overhead.

\section{Related Work}
\label{sec:related_work}

% Existing URLLC studies improve reliability through techniques such as adaptive repetition, reserved transmission opportunities, multiple configured grants, and contention-aware resource allocation \cite{wu2023,le2019,singh2017,mahmood2019,liu2022}. Reactive Hybrid Automatic Repeat reQuest (R-HARQ) is an important reliability mechanism, where retransmissions are triggered after the transmitter receives a negative acknowledgement. Although R-HARQ avoids unnecessary retransmissions, its feedback-dependent operation may limit the number of retransmission attempts available within stringent URLLC deadlines \cite{le2020feedback}.

Existing URLLC communication protocols improve reliability and latency through mechanisms such as adaptive repetition, reserved transmission opportunities, multiple configured grants, contention-aware allocation, grant-free access, and feedback enhancement \cite{wu2023,le2019,singh2017,mahmood2019,liu2022,le2020feedback,goktepe2020}. Among them, reactive HARQ is a classical technique that initiates retransmissions only after receiving negative acknowledgements, thereby avoiding unnecessary transmissions but limiting the number of attempts available within stringent deadlines. \(K\)-Repetition removes this feedback dependency by transmitting a fixed number of consecutive repetitions, reducing retransmission delay at the cost of using all configured opportunities. Proactive HARQ combines preconfigured repetitions with feedback-based early termination, allowing unused opportunities to be released once successful decoding becomes observable.

Formal analysis and scheduling methods have been developed for periodic URLLC traffic under \(K\)-Repetition and other fixed-transmission models \cite{pan2024,zhang2023}. These methods generally represent each packet or flow as a fixed transmission block and construct non-overlapping schedules. Such formulations cannot fully exploit proactive-HARQ early termination because all potential repetitions remain reserved. Probabilistic resource sharing has also been considered in real-time wireless networks \cite{zhang2022reliable,brummet2021warp,deng2017timely}, but existing models commonly abstract transmission and feedback within the same slot and therefore do not capture multi-slot HARQ feedback observability. Ignoring this delay may overestimate the transmission opportunities available for reuse and consequently overestimate the achievable reliability.

The impact of delayed feedback has been recognized in dynamic packet scheduling, where online policies are designed to optimize expected delivery performance or long-term age-related metrics \cite{kim2015optimal,ji2024age}. However, scheduling-signaling configuration and access delays may render such dynamic decisions outdated by the time of transmission, preventing their direct application to 5G systems. In contrast, our work considers offline proactive-HARQ configuration for heterogeneous periodic flows. The proposed DTMC explicitly captures delayed feedback observability and verifies packet-level reliability and deadline requirements under preconfigured resource sharing.

\section{System Model and Problem Statement}

\subsection{Network Model}

We consider a single-cell downlink 5G NR network at the medium access control (MAC) layer, where a base station (BS) sends a set of traffic flows to multiple user equipments (UEs). All flows are treated as independent scheduling entities in this work. The network adopts a set of shared orthogonal channels for downlink transmissions, and time is discretized into equal-length transmission slots.

\subsection{Traffic Model}
Let \( \mathcal{F} = \{f_i \mid 1 \leq i \leq N\} \) denote the set of flows. Each flow generates an infinite sequence of packets that arrive periodically and are transmitted to the UEs under stringent latency and reliability requirements. Specifically, each flow \( f_i \) is characterized by a tuple \( \langle T_i, D_i, P_i \rangle \), where:
\begin{itemize}
    \item \( T_i \): The inter-arrival period of packets in the flow.
    \item \( D_i \): The relative deadline, indicating the maximum delay allowed from packet arrival to successful transmission. We consider constrained deadlines, i.e., \( D_i \leq T_i \).
    \item \( P_i \): The reliability requirement, defined as the minimum probability that each packet must be successfully transmitted within its deadline.
\end{itemize}
For each flow \( f_i \), let \( f_{i,j} \) denote its \(j\)-th packet. To meet reliability requirement \(P_i\), the scheduler allocates sufficient transmission slots within the \(D_i\)-slot window after \(f_{i,j}\) arrives. Each transmission succeeds independently with probability \(p\).\footnote{The transmission success probability can be modeled either as a time-varying value \(p(t)\)~\cite{sadeghi2008finite,baddour2005autoregressive} derived from channel conditions and PHY-layer mechanisms, or as a conservative lower bound \(p\). For simplicity, this paper uses a fixed success probability \(p\) in the following analysis.}

Let \( R \) denote the HARQ round-trip time (RTT) in slots, i.e., the delay from transmission to feedback, which defines the minimum interval between a transmission and its retransmission in reactive HARQ.

To capture interactions among flows with different periods, we define the \emph{hyperperiod} \( H \) as the least common multiple of all periods:
\[
H = \text{lcm}(T_1, T_2, \ldots, T_N).
\]
Within one hyperperiod, flow \( f_i \) generates \( H / T_i \) packets, and the complete set of packets is
\[
\{ f_{i,j} \mid 1 \leq i \leq N,\ 1 \leq j \leq H / T_i \}.
\]

\subsection{Scheduling Model}
\label{Scheduling}

We consider a grant-free downlink model where strictly periodic flows share preconfigured resources without dynamic scheduling overhead.

Scheduling is performed at the flow level by configuring transmission patterns. For each flow \( f_i \), a release offset \( o_i \in [0, T_i) \) is assigned to determine packet arrival times. Each flow is also associated with parameters \( \{K_{i,1}, K_{i,2}, \ldots\} \), where \( K_{i,j} \) denotes the number of pre-allocated transmission slots for packet \( f_{i,j} \).

Under this setting, packet \(f_{i,j}\) arrives at slot \(o_i+(j-1)T_i\). Packets are served sequentially according to their arrival times, meaning that a packet can use its allocated slots only after all preceding packets complete. Once eligible, it performs consecutive transmissions until successful decoding or all opportunities are exhausted. Since flows reuse the same resource, their configured transmission slots may overlap, but only one packet can be actually transmitted in each slot.

The scheduling problem is to determine the offsets and slot allocations such that all latency and reliability requirements are satisfied. Formally, the configuration of flow \(f_i\) is
\[
c_i=\left\langle o_i,\{K_{i,1},K_{i,2},\ldots,K_{i,n}\}\right\rangle .
\]

\subsection{HARQ Models}

Reliability is typically achieved through redundant transmissions. In the absence of interference from other flows, the minimum number of transmission attempts required to meet the reliability requirement \( P_i \) satisfies

\[
1 - (1 - p)^{K_i} \geq P_i.
\]
Equivalently,
\begin{equation}
K_i \geq \left\lceil \frac{\ln(1 - P_i)}{\ln(1 - p)} \right\rceil.
\label{equ1}
\end{equation}

\begin{figure}[!t]
    \centering
    \includegraphics[width=0.9\linewidth]{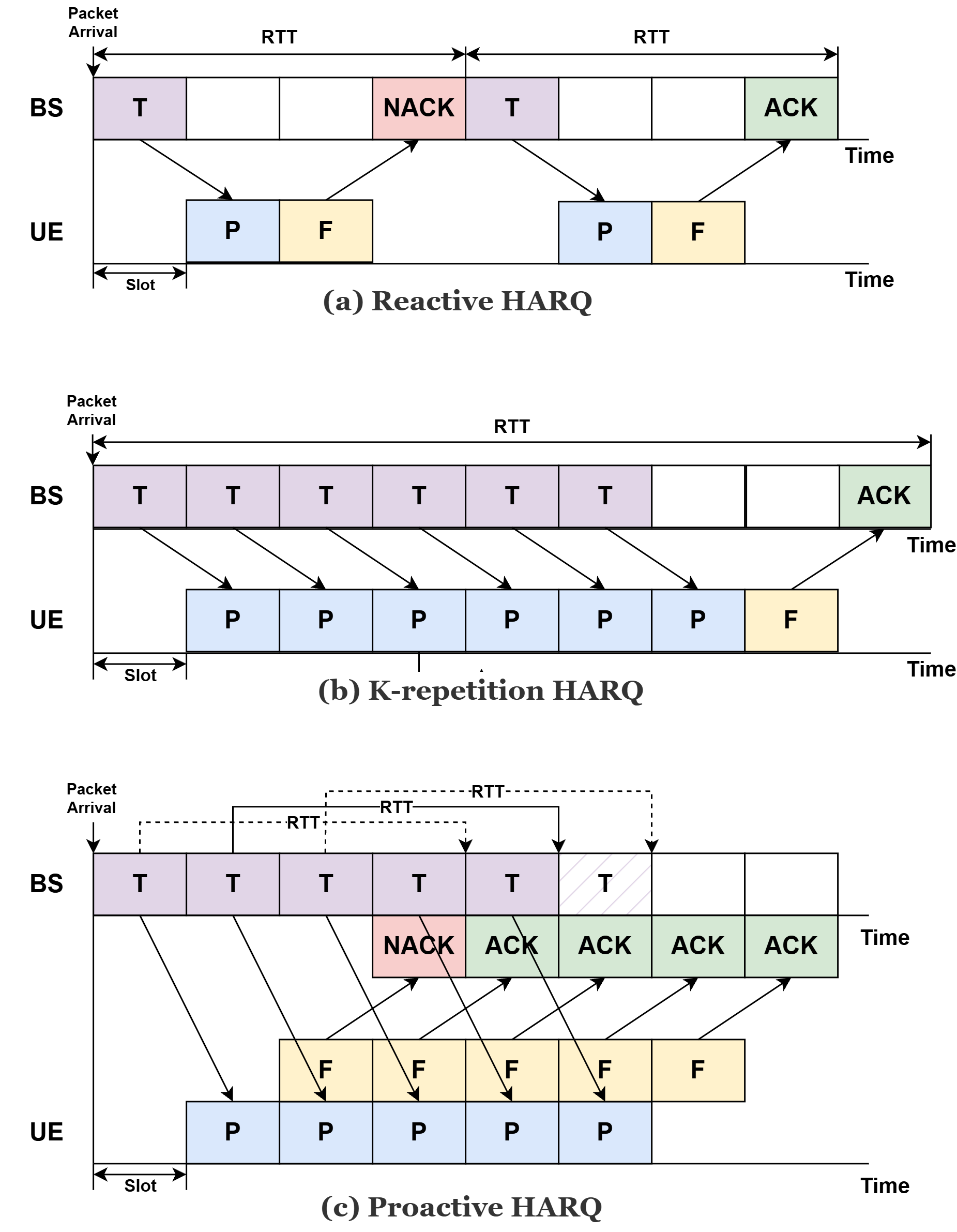}
    \caption{Illustration of HARQ mechanisms: Reactive HARQ, $K$-Repetition, and Proactive HARQ ($RTT=4$). Legend: T = transmission, P = processing, F = feedback, ACK = reception.}
    \vspace{-1.5em}
    \label{fig:harq_mechanisms}
\end{figure}

As introduced in Section~\ref{sec:related_work} and illustrated in Fig.~\ref{fig:harq_mechanisms}, redundancy can be realized through reactive HARQ, \(K\)-repetition, or proactive HARQ. A HARQ process includes transmission, reception and processing, feedback transmission, and feedback reception, resulting in an RTT of four slots. Reactive HARQ incurs a full RTT between attempts, whereas \(K\)-repetition postpones feedback until all repetitions finish. Proactive HARQ instead supports early termination after successful decoding, providing a better balance between latency and resource efficiency.

% \textbf{Reactive HARQ:} Retransmissions are triggered only upon NACK feedback, which avoids unnecessary transmissions but incurs additional latency due to the feedback delay.

% \textbf{K-repetition:} The transmitter sends $K$ consecutive replicas of a packet over pre-allocated slots without waiting for feedback. The receiver performs decoding after all replicas are received and sends a single ACK/NACK. No further transmissions are performed regardless of the outcome.

% \textbf{Proactive HARQ:} Similar to $K$-repetition, but feedback is provided after each transmission attempt. The transmitter stops immediately upon receiving an ACK, even if pre-allocated slots remain unused.

\subsection{Problem Formulation}
\label{subsection:Problem Formulation}

In URLLC communications, the key objective is to ensure that each packet is successfully delivered within its deadline while satisfying its target reliability. Specifically, each packet \(f_{i,j}\) of flow \(f_i\) must be successfully delivered within
\[
[\,o_i + (j-1)T_i,\; o_i + (j-1)T_i + D_i\,]
\]
with probability at least \(P_i\), thereby satisfying the relative deadline \(D_i\).

As discussed in Section~\ref{Scheduling}, the scheduling problem is to determine the release offsets \(o_i\) and the transmission counts \(K_{i,j}\). The offsets determine how packets from different flows share configured resources, while \(K_{i,j}\) determines the tradeoff between reliability and resource usage: too few opportunities may violate reliability, whereas too many opportunities may waste resources or cause deadline misses. Different retransmission mechanisms impose different constraints on \(K_{i,j}\). 

\subsubsection{Reactive HARQ}
\label{sec:reactive HARQ}
Under reactive HARQ, two consecutive attempts of the same packet must be separated by at least \(R\) slots in terms of their starting slots. Thus, without interference, feasibility requires
\begin{equation}
\left\lceil \frac{\ln(1-P_i)}{\ln(1-p)} \right\rceil 
\le K_{i,j} 
\le 1+\left\lfloor \frac{D_i-1}{R} \right\rfloor .
\label{equ2}
\end{equation}
This limits the number of retransmissions available under stringent URLLC deadlines.

\subsubsection{\(K\)-Repetition}
\label{sec:Repetition}

Under \(K\)-repetition, transmissions are consecutive and feedback is not used for early termination; therefore, a non-overlapping configuration is required for every flow pair \(i,j\). The repetition number \(K_i\), obtained from Eq.~\eqref{equ1} to satisfy reliability, must also satisfy \(K_i\le D_i\). Similar to~\cite{zhang2023}, a contention-free configuration can be obtained using the condition in~\cite{aarts1991parle}:
\begin{equation}
K_i
\le
(o_j-o_i)\bmod\gcd(T_i,T_j)
\le
\gcd(T_i,T_j)-K_j.
\label{eq:static-conflict}
\end{equation}
Since all configured repetitions are transmitted, resources may be wasted after reliability has already been achieved.

\subsubsection{Proactive HARQ}
\label{sec:proactive HARQ}
Proactive HARQ combines proactive transmission opportunities with feedback-based early termination. It can release unused opportunities after successful decoding, but this also makes the design of \(K_{i,j}\) more challenging: the actual resource usage depends on stochastic transmission outcomes and delayed feedback. In this paper, we model this process using a DTMC and derive transmission configurations that satisfy both reliability and deadline constraints.

\section{Discrete-Time Markov Chain-Based Analysis}

Under proactive HARQ, one can directly use the same preconfigured transmission allocation as \(K\)-repetition, which guarantees reliability if the corresponding schedule is feasible. However, this ignores early termination and still reserves all potential repetitions, leading to unnecessary resource waste. 

In proactive HARQ, the actual number of transmissions per packet is stochastic, since remaining opportunities are terminated after successful decoding. To capture this behavior, we model the transmission process as a discrete-time Markov chain and analyze packet reliability across transmission slots.

\subsection{Discrete-Time Markov Chain}
A Discrete-Time Markov Chain (DTMC) is a stochastic process that evolves over a discrete time index \( t = 0, 1, 2, \ldots \), and is characterized by the \emph{Markov property}, which states that the future state depends only on the current state and not on the sequence of previous states.

Formally, let \(\{X_t\}_{t\in\mathbb{N}}\) be a stochastic process over a finite or countable state space \(\mathcal{S}=\{s_1,s_2,\ldots,s_n\}\). It is a DTMC if, for all \(t\in\mathbb{N}\) and \(s_i,s_j\in\mathcal{S}\),
\[
\Pr(X_{t+1} = s_j \mid X_t = s_i, \ldots, X_0 = x_0) \\
= \Pr(X_{t+1} = s_j \mid X_t = s_i)
\]

The chain is characterized by a transition probability matrix \( \mathbf{P} \in \mathbb{R}^{n \times n} \), where each entry
\[
P_{i,j} = \Pr(X_{t+1} = s_j \mid X_t = s_i), \quad \sum_{j=1}^{n} P_{i,j} = 1,
\]
represents the one-step transition probability from state \( s_i \) to state \( s_j \).

The initial distribution over the states is denoted by a row vector \( \boldsymbol{\pi}^{(0)} = [\pi^{(0)}_1, \pi^{(0)}_2, \ldots, \pi^{(0)}_n] \), where \( \pi^{(0)}_i = \Pr(X_0 = s_i) \). The distribution after \( t \) time steps is given by:
\[
\boldsymbol{\pi}^{(t)} = \boldsymbol{\pi}^{(0)} \mathbf{P}^t.
\]

A state \( s_i \in \mathcal{S} \) is called an \emph{absorbing state} if, once entered, the process remains in that state forever. This is mathematically defined as:
\[
P_{i,i} = 1 \quad \text{and} \quad P_{i,j} = 0 \quad \text{for all } j \neq i.
\]
A DTMC is said to be an \emph{absorbing Markov chain} if it has at least one absorbing state and every non-absorbing state can eventually reach an absorbing state with positive probability.

\subsection{Continuous Transmission}
\label{sec:Continuous}

We consider a scenario in which multiple packets share the same configured resource and are served sequentially over the allocated transmission slots. This setup can be modeled using a Discrete-Time Markov chain.

Let the state \( s_n \), where \( n \in \{0, 1, \ldots, N\} \), denote the state in which exactly \( n \) packets have been successfully delivered. In particular, state \( s_0 \) represents that no packets have been successfully transmitted, while state \( s_N \) is an absorbing state indicating that all packets have been successfully delivered.

Since each independent transmission has a success probability of \( p \), the DTMC transitions from state \( s_n \) to state \( s_{n+1} \) with probability \( p \), or remains in state \( s_n \) with probability \( 1 - p \).
The corresponding transition matrix \( \mathbf{P} \in \mathbb{R}^{(N+1) \times (N+1)} \) for the DTMC model is given by:

\[
\mathbf{P} =
\begin{bmatrix}
1 - p & p      & 0      & \cdots & 0      & 0 \\
0     & 1 - p  & p      & \cdots & 0      & 0 \\
0     & 0      & 1 - p  & \cdots & 0      & 0 \\
\vdots & \vdots & \vdots & \ddots & \vdots & \vdots \\
0     & 0      & 0      & \cdots & 1 - p  & p \\
0     & 0      & 0      & \cdots & 0      & 1
\end{bmatrix}
\]

Let \( \boldsymbol{\pi}^{(k)} = [\pi^{(k)}_0, \pi^{(k)}_1, \ldots, \pi^{(k)}_N] \) denote the state distribution after \( k \) transmissions, where \( \pi^{(k)}_n \) represents the probability that the system is in state \( s_n \) at step \( k \). Assuming the initial distribution is \( \boldsymbol{\pi}^{(0)} = [1, 0, \ldots, 0] \), which evolves as
\begin{equation}
\boldsymbol{\pi}^{(k)} = \boldsymbol{\pi}^{(0)} \mathbf{P}^k.
\label{equ4}
\end{equation}

To ensure the reliability requirement of packet \( f_{i,j} \), which corresponds to reaching state \( s_q \), we seek the minimum number of transmission opportunities \( k \) such that
\begin{equation}
    \Pr[X_k \geq q] = \sum_{n = q}^{N} \pi^{(k)}_n \geq P_i.
    \label{equ5}
\end{equation}

% where \( \pi^{(k)}_n \) denotes the probability that the system is in state \( s_n \) after \( k \) transmission opportunities.

Under the sequential transmission model, this tail probability corresponds exactly to the successful delivery probability of packet \( f_{i,j} \), since the transmission of this packet implies that all preceding packets have also been successfully delivered.

That is,
\begin{equation}
k^* = \min \left\{ k \in \mathbb{N} \;\middle|\; \sum_{n = q}^{N} \pi^{(k)}_n \geq P_i \right\}.
\end{equation}

In this setting, the configuration \( c_{i} \) is extended by adding the parameter \(K_{i,j}=k^*\), meaning that packet \(f_{i,j}\) can utilize at most \(k^*\) transmission slots. If the reliability requirement cannot be satisfied within the available slots, e.g., due to deadline constraints, the analysis terminates and the corresponding flow is considered unschedulable.

Once a packet meets its reliability requirement, it is regarded as delivered in subsequent analysis. Accordingly, after \( k^* \) transmission slots, the state distribution \( \boldsymbol{\pi}^{(k^*)} \) is updated by merging the probability mass of \( s_0 \) into \( s_1 \), removing \( s_0 \), and treating \( s_1 \) as the new initial state:
\[
\boldsymbol{\pi}^{(k^*)}_{\text{new}} =
\left[
\pi^{(k^*)}_1 + \pi^{(k^*)}_0,\;
\pi^{(k^*)}_2,\;
\ldots,\;
\pi^{(k^*)}_N
\right].
\]
This operation reflects a conditional progression in which the first packet is assumed completed and is no longer tracked.

\begin{figure}[!t]
\centering
\includegraphics[width=\linewidth]{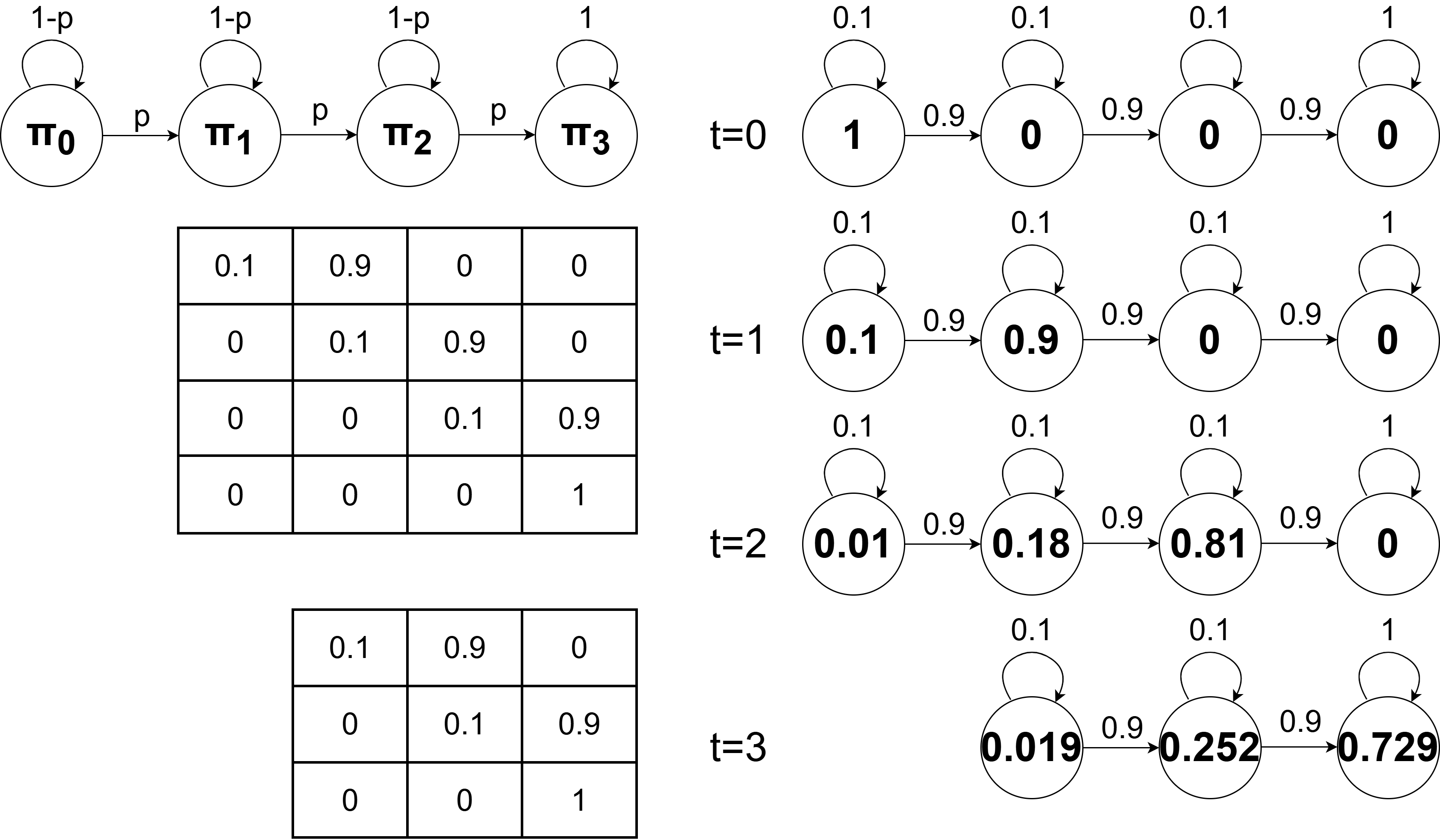}
\caption{An example of the Markov chain for packet transmission and the corresponding transformation of the state transition matrix.}
\vspace{-1.5em}
\label{fig:continue}
\end{figure}

Figure~\ref{fig:continue} shows the state variation when the single-transmission success probability is $0.9$ and each packet is required to achieve a reliability of $99\%$. 
From $t=0$ to $t=2$, one transmission attempt is performed in every slot, which is mathematically represented by multiplying the state vector with the state transition matrix shown on the left.

At \(t=2\), the cumulative probability of states \(\pi_1\) and beyond reaches \(99\%\), meaning that the first packet meets its required reliability.
Therefore, from the next slot onward, this packet is no longer transmitted.
The probability mass of \(\pi_0\) is then added to \(\pi_1\) to fold the state vector, and the new state vector is multiplied by the transformed transition matrix shown at the lower left to obtain the state at \(t=3\).
In this folded representation, \(\pi_1\) becomes the new initial state.

\subsection{Impact of HARQ Round-Trip Time}
\label{subsection:HARQ Round-Trip Time}

In practice, due to hardware and processing delays, the HARQ protocol incurs a round-trip time of \( R \) slots. Specifically, a transmission initiated at slot \( t \) results in an ACK/NACK that becomes available at the end of slot \( t + R - 1 \), and can only be utilized by the transmitter starting from slot \( t + R \). Consequently, proactive HARQ can terminate ongoing repetitions no earlier than slot \( t + R \).

To incorporate the HARQ RTT into the DTMC, we expand each logical “success state” into multiple substates that capture the delay in feedback. Specifically, for each state \( s_n \), we introduce \( R \) substates
\[
s_{n,1}, s_{n,2}, \ldots, s_{n,R},
\]
where \( s_{n,t} \) denotes that the system is in success state \( s_n \) and at the \( t \)-th slot of the \( R \)-slot HARQ feedback cycle. The transition rules are defined as follows:
\begin{itemize}
  \item For \(t = 1, \dots, R-1\): there is a deterministic advance within the HARQ cycle, indicating that the transmitter has not yet received the ACK corresponding to state \(s_n\) and continues transmitting,
  \[
    (n,t) \longrightarrow (n,t+1) \quad \text{with probability } 1.
  \]

  \item At the feedback slot \(t=R\): the ACK for the transmission that started \(R-1\) slots earlier arrives. At this point, the next packet can start transmission with probability \(p\); if successful, the system transitions to the next state
  \[
    (n,R) \longrightarrow (n+1,1) \quad \text{with probability } p,
  \]
  otherwise, with probability \(1-p\), the system remains in the current substate,
  \[
    (i,R) \longrightarrow (i,R) \quad \text{with probability } 1-p,
  \]

  \item If \(n = N\), corresponding to the final packet, the state is absorbing.
\end{itemize}

The corresponding transition matrix \(\mathbf{P}_{\text{single}}\) for a packet is:
\[
\mathbf{P}_{\text{single}} =
\begin{bmatrix}
1-p & p & 0 & \cdots & 0\\
0 & 0 & 1 & \cdots & 0\\
0 & 0 & 0 & \ddots & 0\\
\vdots & \vdots & \vdots & \ddots & 1\\
0 & 0 & 0 & \cdots & 1
\end{bmatrix},
\]
where the matrix is of size \( (R+1) \times (R+1) \), including the initial state \( s_0 \) and the substates associated with \( s_1 \).

The first row describes the transmission outcome from \(s_0\): with probability \(p\), the packet is successfully transmitted and the system moves to \(s_{1,1}\); with probability \(1-p\), it remains in \(s_0\). The remaining rows describe the deterministic progression through the feedback-delay substates. Thus, \(s_0\) has no substates and only transitions to itself or to \(s_{1,1}\).

To model \( N \) packets transmitted consecutively, we replicate and expand the single-packet state space for each packet, inserting \( R-1 \) waiting states between consecutive success states. The resulting  transition matrix has dimension \( N \cdot R + 1 \).
\[
\mathbf{P}_{\text{HARQ}} = 
\begin{bmatrix} 
1-p & p & 0 & \cdots & 0 & 0 & \cdots & 0 \\
0 & 0 & 1 & \cdots & 0 & 0 & \cdots & 0 \\
\vdots & \vdots & \vdots & \ddots & \vdots & \vdots & & \vdots \\
0 & 0 & 0 & \cdots & 1 & 0 & \cdots & 0 \\
0 & 0 & 0 & \cdots & 1-p & p & \cdots & 0 \\
0 & 0 & 0 & \cdots & 0 & 0 & 1 & \vdots \\
\vdots & \vdots & \vdots & \vdots & \vdots & \vdots & \ddots & \vdots \\
0 & 0 & 0 & \cdots & 0 & 0 & \cdots & 1 \end{bmatrix}.
\]

The first row of \( \mathbf{P}_{\text{HARQ}} \) corresponds to the initial state \( s_0 \), and the following rows represent substates \( s_{1,1}, \ldots, s_{1,R} \) of the first packet. State \( s_{1,1} \) transitions deterministically through \( s_{1,2} \), \( s_{1,3} \), and \( s_{1,R} \). At \( s_{1,R} \), the system moves to \( s_{2,1} \) with probability \( p \) or remains in \( s_{1,R} \) with probability \( 1-p \).
The last row corresponds to state \( s_{N,R} \) of the final packet, which is absorbing as no further transitions occur after receiving ACK.

To incorporate the HARQ RTT, the system state distribution vector is expanded to include the intermediate waiting states for each packet. The updated distribution is denoted as:
\[
\boldsymbol{\pi}_{\text{RTT}} = [\, \pi_0, \pi_{1,1}, \dots, \pi_{1,R}, \pi_{2,1}, \dots, \pi_{N,R} \,],
\]

where each packet is represented by \(R\) consecutive RTT states. Similarly to Eq.~\eqref{equ4}, the initial state vector is given by
\(
\boldsymbol{\pi}^{(0)}_{\text{RTT}} = [1, 0, \ldots, 0],
\)
which evolves as:
\begin{equation}
\boldsymbol{\pi}^{(k)}_{\text{RTT}} = \boldsymbol{\pi}^{(0)}_{\text{RTT}} \, \mathbf{P}_{\text{HARQ}}^k.
\label{equ7}
\end{equation}

Based on Eq.~\eqref{equ5}, the reliability requirement \(P_i\) for transmitting at least \(q\) packets can be evaluated using the expanded state distribution \(\boldsymbol{\pi}^{(k)}_{\text{RTT}}\), which is:
\begin{equation}
\label{equ8}
\Pr[X_k \geq q] =  \sum_{n = q}^{N}\sum_{t = 1}^{R} \pi^{(k)}_{n,t} \geq P_{\text{i}}.
\end{equation}

In the special case of \(q=1\), this simplifies to
\[
\Pr[X_k \geq 1] = 1 - \pi^{(k)}_0,
\]

Similar to Section~\ref{sec:Continuous}, when computing \( \boldsymbol{\pi}^{(k)}_{\text{RTT}} \), once the reliability requirement of the packet associated with state \(s_n\) is satisfied, we perform state aggregation. Specifically, the preceding state and all substates of packet \(n\) are merged into a new aggregated state.

Formally, suppose the current state distribution is
\[
\boldsymbol{\pi}^{(k)}_{\text{old}} =
\Big[
\pi^{(k)}_{n-1}, \;
\pi^{(k)}_{n,1}, \pi^{(k)}_{n,2}, \;
\ldots, \pi^{(k)}_{N,R}
\Big].
\]

If at this step the reliability condition is satisfied, i.e.,
\[
1 - \pi^{(k)}_{n-1} \geq P_i,
\]
we update the state probability by aggregating
\[
\pi^{(k)}_{n} \leftarrow \pi^{(k)}_{n-1} + \sum_{t=1}^{R} \pi^{(k)}_{n,t}.
\]

The updated state distribution becomes
\[
\boldsymbol{\pi}^{(k)}_{\text{new}} =
\Big[
\pi^{(k)}_{n}, \;
\pi^{(k)}_{n+1,1}, \pi^{(k)}_{n+1,2}, \;
\ldots, \pi^{(k)}_{N,R}
\Big].
\]

% This operation reflects that after \( R \) transmission slots, the probability of successfully delivering packet \( i \) exceeds the required reliability threshold. Therefore, packet \( i \) (as well as all preceding packets) is considered successfully delivered and is no longer tracked. As a result, the probability mass of all corresponding substates is accumulated into state \( s_i \).

% Accordingly, the transition matrix is updated by removing the transitions associated with the aggregated states and incorporating the transitions of the newly formed state.

% Using this procedure, we can determine the effective repetition opportunities \(R_{i,j}\) for each packet, i.e., the number of transmission slots that the packet can actually use between its arrival and the folding time \(k\). This ensures that, despite potential contention, the reliability requirement \(P_i\) is met.

Using this procedure, we allocate transmission opportunities for each packet. For the first packet, the required number of transmission slots is
\begin{equation}
K_1 = \min \left\{ k \in \mathbb{N} \;\middle|\; 
\sum_{n=1}^{N} \sum_{t=1}^{R} \pi_{n,t}^{(k)} \geq P_1
\right\}.
\end{equation}

After \(K_1\) steps, the state vector and transition matrix are updated through state aggregation. By iterating this procedure, we determine the required transmission opportunities for each packet in order. For the \(x\)-th packet, this is given by
\begin{equation}
K_x = \min \left\{ k \in \mathbb{N} \;\middle|\; 
\sum_{n=x}^{N} \sum_{t=1}^{R} \pi_{n,t}^{(k)} \geq P_x
\right\}.
\end{equation}

% \begin{figure}[!t]
% \centering
% \includegraphics[width=\linewidth]{markov.drawio.png}
% \caption{An example of the folded Markov chain for HARQ-based packet transmission and the corresponding transformation of the state transition matrix.}
% \label{fig:markov}
% \end{figure}

% Figure~\ref{fig:markov} illustrates a Markov chain for a HARQ-enabled transmission process with $K=2$ and a per-attempt success probability $p$. 
% The states $\pi_{1,1}$ and $\pi_{1,2}$ represent all states of the first packet, corresponding to the first slot and the second slot after a success occurs, respectively. 
% When $K=2$, an ACK feedback is received only in the second slot after a successful transmission, at which point the UE is allowed to start transmitting the second packet.

% Once the success probability of the packet corresponding to $\pi_1$ satisfies the reliability requirement, the transmitter stops sending this packet in the next slot regardless of the transmission outcome, and the system always proceeds to transmit the packet corresponding to $\pi_2$. 
% Consequently, the next states of $\pi_0$, $\pi_{1,1}$, and $\pi_{1,2}$ all become the transmission state of the second packet. 
% To reduce the number of maintained states, the probabilities of these states are aggregated into a new $\pi_0$ state, and the state $\pi_2$ is relabeled as the new $\pi_1$. 
% When more than two packets are waiting for transmission, the same state aggregation and relabeling procedure is applied iteratively to subsequent packets.

\subsection{Additional Applications of State Expansion}
\label{subsection:state_extension}

The state-expansion principle can also model other fine-grained timing effects in 5G NR. Deterministic intermediate states represent timing costs, while state-dependent transition probabilities capture reliability gains.

\subsubsection{Non-Unit Transmission Duration}

In 5G NR, if one transmission attempt spans multiple scheduling units, its BLER-based success probability is applied only after completion. For example, if one transmission occupies two units, the state space \([s_0,s_{\mathrm{tx}},s_1]\) gives
\[
\mathbf{P}_{\mathrm{exec}} =
\begin{bmatrix}
0 & 1 & 0\\
1-p & 0 & p\\
0 & 0 & 1
\end{bmatrix},
\]
where \(s_{\mathrm{tx}}\) denotes an ongoing transmission. Longer transmissions can be modeled by more deterministic states.

For minislot-based repetitions~\cite{gerami2023configured}, as discussed in~\cite{le2020feedback}, symbol-level constraints may cause multiple transmissions within a slot to share one HARQ feedback. If \(k\) minislot repetitions are performed within one slot and each succeeds with probability \(p\), the effective slot success probability is
\[
p_{\mathrm{slot}} = 1-(1-p)^k .
\]

\subsubsection{Soft Combining}

When redundancy versions are jointly decoded, the transition probability should reflect accumulated decoding gain\cite{ding2021optimized}. For at most two combined transmissions, let \(p_1\) and \(p_2\) be the success probabilities after the first and second transmissions with soft combining given that the first fails, respectively. 
Then
\[
\mathbf{P}_{\mathrm{sc}} =
\begin{bmatrix}
0 & 1-p_1 & p_1 \\
0 & 1-p_2 & p_2 \\
0 & 0 & 1
\end{bmatrix}.
\]

This models reliability gains from accumulated redundancy versions using state-dependent transition probabilities.

\subsubsection{Numerology-Dependent Timing}

In 5G NR, different numerologies correspond to different slot durations,
and systems may use different numerologies at different scheduling intervals~\cite{esmaeily2023beyond,bag2019multi,sexton2019customization}. 
When the scheduling unit changes, the HARQ feedback delay must be re-expressed in the new unit. Let \(T_{\mathrm{old}}\) and \(T_{\mathrm{new}}\) be the old and new unit durations. If the absolute feedback delay is unchanged, we have 
\(
R_{\mathrm{new}}
=
\left\lceil
\frac{R_{\mathrm{old}}T_{\mathrm{old}}}{T_{\mathrm{new}}}
\right\rceil .
\)

A shorter unit expands each waiting state into finer substates, whereas a longer unit merges multiple substates into one coarser substate. If \(k=T_{\mathrm{new}}/T_{\mathrm{old}}\in\mathbb{N}\), old waiting substate \(t\) is conservatively aligned to
\[
h(t)=1+k\left\lceil \frac{t-1}{k}\right\rceil ,
\]
and probability masses mapped to the same boundary are aggregated:
\[
\bar{\pi}_{n,\tau}
=
\sum_{t:\,h(t)=\tau}
\pi_{n,t}
\qquad
\tau \in \{1,1+k,1+2k,\ldots\}.
\]

This prevents using feedback that arrives between two coarse scheduling boundaries before the next valid boundary.

\section{Schedulability Test and Schedule Table}
\label{sec:algorithm}

% In this section, we apply the proposed analytical framework to analyze the schedulability of a given flow set under proactive HARQ. Using the method introduced in Section~\ref{subsection:HARQ Round-Trip Time}, we first check whether all flows can satisfy their latency and reliability requirements in the worst case. If not, we then construct a feasible schedule table by configuring transmission offsets and repetition parameters.

In this section, we first present how to analyze the schedulability of a flow set under a given release-offset configuration. Based on this analysis, we then develop a two-stage genetic algorithm that searches for feasible offset configurations.

\subsection{DTMC-Based Schedulability Test}

A flow set is schedulable under a given offset configuration if every packet satisfies both its latency deadline and reliability requirement. To improve the likelihood of meeting latency constraints, each packet should begin transmission as early as possible in its arrival slot. We next describe how the flow set is analyzed slot by slot under the given release offsets.

At the beginning of the analysis, the state vector and transition matrix are initialized as
\begin{equation}
\boldsymbol{\pi}^{(0)}=[1], \quad \mathbf{P}^{(0)}=[1],
\end{equation}
respectively. At this stage, the system contains only state \(s_0\), which indicates that no packet has been successfully transmitted, and all probability mass is assigned to this state.

\begin{figure}[!t]
\centering
\includegraphics[width=\linewidth]{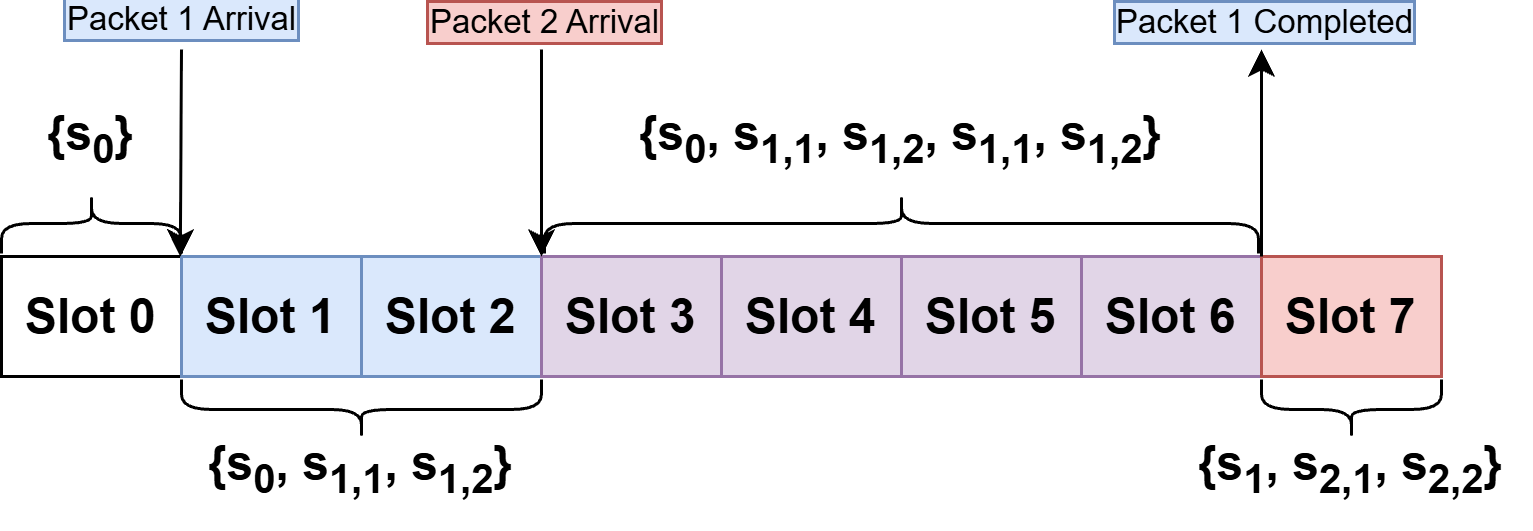}
\caption{Expansion of the state space upon packet arrival and completion.}
\vspace{-1em}
\label{fig:Packet-Arrival}
\end{figure}

\begin{figure}[!t]
\centering
\includegraphics[width=\linewidth]{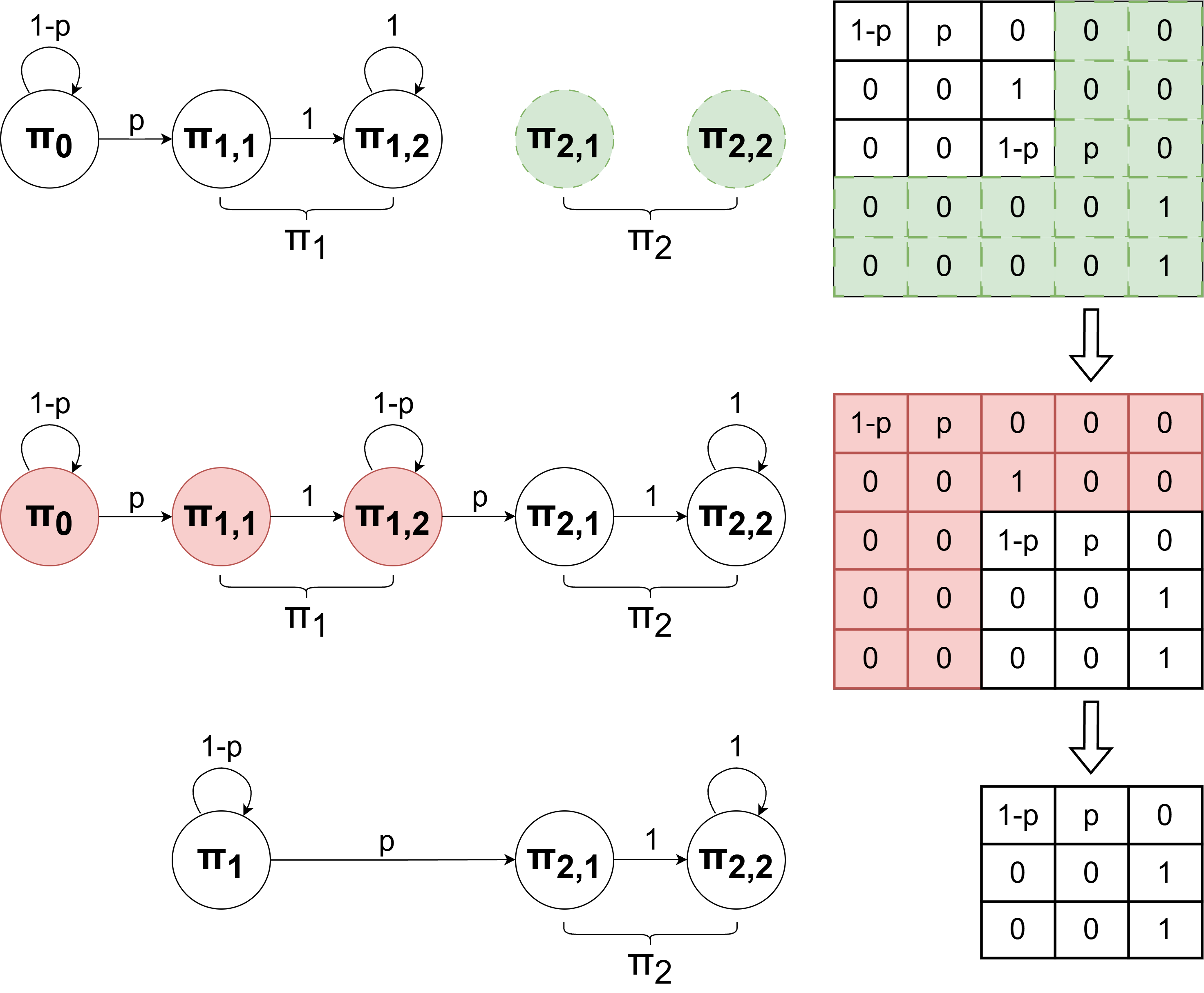}
\caption{Illustration of packet arrivals and completions and the corresponding evolution of the transition matrix.}
\vspace{-1em}
\label{fig:markov}
\end{figure}

When a new packet arrives, it is appended to the packet queue according to the FIFO policy. As shown in Fig.~\ref{fig:Packet-Arrival}, its \(R\) HARQ states are appended to the existing state space. Accordingly, as illustrated in Fig.~\ref{fig:markov}, the state vector is expanded by appending \(R\) zero-valued elements, while the transition matrix is expanded by adding the corresponding rows and columns, as described in Section~\ref{subsection:HARQ Round-Trip Time}. The existing probability mass and transitions remain unchanged.

Let \(\mathbf{P}^{(t)}\) denote the transition matrix for slot \(t\), the state distribution is updated as
\begin{equation}
\boldsymbol{\pi}^{(t+1)}
=
\boldsymbol{\pi}^{(t)}\mathbf{P}^{(t)}.
\end{equation}

After each update, the reliability and deadline conditions of the packets are checked using Eq.~\eqref{equ8}. Once the first packet satisfies its reliability requirement, it is regarded as completed and the initial state is updated to \(s_1\). Accordingly, \(s_0\) and the substates of \(s_1\) are removed, and their probability masses, \(\pi_0\) and \(\pi_{1,t}\), are aggregated into the new initial-state probability \(\pi_1\), and the transition matrix is reduced by \(R\) dimensions. 

The analysis continues until all packets complete or a packet fails to satisfy its reliability requirement by its deadline. Therefore, for any given offset configuration, the proposed DTMC-based test determines whether every packet released within one hyperperiod satisfies its reliability and deadline requirements. The flow set is declared schedulable under that configuration only if all packets pass the test.

% As illustrated in Fig.~\ref{fig:markov}, the system initially consists of states \(s_0\), \(s_{1,1}\), and \(s_{1,2}\) (where $\pi_i$ denotes the probability mass of state $s_i$). When the second packet arrives, state \( s_{1,2} \) transitions from an absorbing state to a transient state preceding \( s_{2,1} \), and subsequently transitions to \( s_{2,1} \) with probability \( p \) each step. The transition matrix then expands from \( 3 \times 3 \) to \( 5 \times 5 \).
% After several iterations, once the reliability condition \(1-\pi^{(k)}_0\ge P_1\) is satisfied, the states \( s_0, s_{1,1}, s_{1,2} \) are aggregated into a new initial state \( s_1 \), and the corresponding transmission parameter \( K_1 \) is determined. The transition matrix then shrinks from \( 5 \times 5 \) back to \( 3 \times 3 \). The updated system consists of states \( s_1, s_{2,1}, s_{2,2} \), and the DTMC evolves according to the reduced transition matrix.

\subsection{Genetic Algorithm and Schedule Table Construction}
\label{sec:alg2}

The preceding DTMC-based test evaluates schedulability for a given offset configuration. For flow sets with configurable offsets, finding a feasible configuration is NP-hard because release offsets and required transmission opportunities are tightly coupled~\cite{zhang2022reliable}. We therefore propose a two-stage algorithm. It first tests the synchronous-release configuration, where \(o_i=0\) for all flows. If feasible, this configuration is used directly; otherwise, a genetic algorithm searches for alternative offsets, with each candidate evaluated by the same DTMC-based test.

\begin{algorithm}[!t]
\caption{Two-Stage Genetic Algorithm for Offset Configuration under Proactive HARQ}
\label{alg:GA-HARQ}
\KwIn{Flow set $\mathcal{F}$, population size $N$, number of generations $G$, elite size $M$}
\KwOut{A feasible offset configuration $\boldsymbol{o}$ and repetition schedule $\mathcal{K}$, or \textit{failure}}

Initialize the configuration $\boldsymbol{o}^{(0)}=\boldsymbol{0}$\;

Perform the DTMC-based analysis over hyperperiod\;

\If{all packets satisfy their latency and reliability requirements}{
    Obtain $\mathcal{K}=\{K_{i,j}\}$\;
    \Return{$(\boldsymbol{o}^{(0)},\mathcal{K})$}\;
}

Initialize a population $\mathcal{P}_0$ of $N$ candidate offset configurations by randomly perturbing $\boldsymbol{o}^{(0)}$\;

\For{$g=0$ \KwTo $G-1$}{
    Compute the static conflict count for each candidate configuration in $\mathcal{P}_g$\;
    
    Select the $M$ configurations with the smallest static conflict counts to form the elite set $\mathcal{E}_g$\;
    
    \ForEach{offset configuration $\boldsymbol{o}\in\mathcal{E}_g$}{
        Perform the DTMC-based schedulability test over one hyperperiod\;
        Obtain the repetition schedule $\mathcal{K}=\{K_{i,j}\}$\;
        
        \If{all packets satisfy their latency and reliability requirements}{
            \Return{$(\boldsymbol{o},\mathcal{K})$}\;
        }
    }
    
    Generate a new population $\mathcal{P}_{g+1}$ through tournament selection, crossover, and mutation\;
}

\Return{failure}\;
\end{algorithm}

As shown in Algorithm~\ref{alg:GA-HARQ}, the first stage tests the synchronous-release configuration because it can be evaluated rapidly. Passing this test directly yields a feasible configuration and avoids unnecessary GA overhead for simple flow sets.

The second stage searches for alternative offsets. The initial population consists of randomized perturbations of the synchronous-release configuration. For each candidate, we compute the total number of flow pairs violating Eq.~\eqref{eq:static-conflict}, referred to as the \emph{static conflict count}.

The \(M\) candidates with the fewest static conflicts are selected for DTMC-based schedulability testing. For each candidate, all packet arrivals within one hyperperiod are enumerated and ordered by arrival time. Packets arriving in the same slot are further ordered by increasing deadlines. Following the procedure in the previous subsection, the state vector and transition matrix are updated as packets arrive and complete, and the state distribution is evaluated slot by slot.

If every packet satisfies its latency and reliability requirements, the corresponding offset configuration and repetition schedule are returned. Otherwise, a new population is generated using tournament selection, crossover, and mutation. This process continues until a feasible configuration is found or the generation limit is reached.

The GA is motivated by two observations. First, fewer transmission conflicts generally provide packets with more dedicated transmission opportunities before their deadlines. We therefore prioritize low-conflict candidates during the search. Second, the static conflict relation between two flows depends only on their offsets, as indicated byEq.~\eqref{eq:static-conflict}. Consequently, non-conflicting offset combinations can be inherited by offspring, allowing the genetic operators to preserve favorable local structures and accelerate convergence.

\begin{figure}[!t] \centering \includegraphics[width=\linewidth]{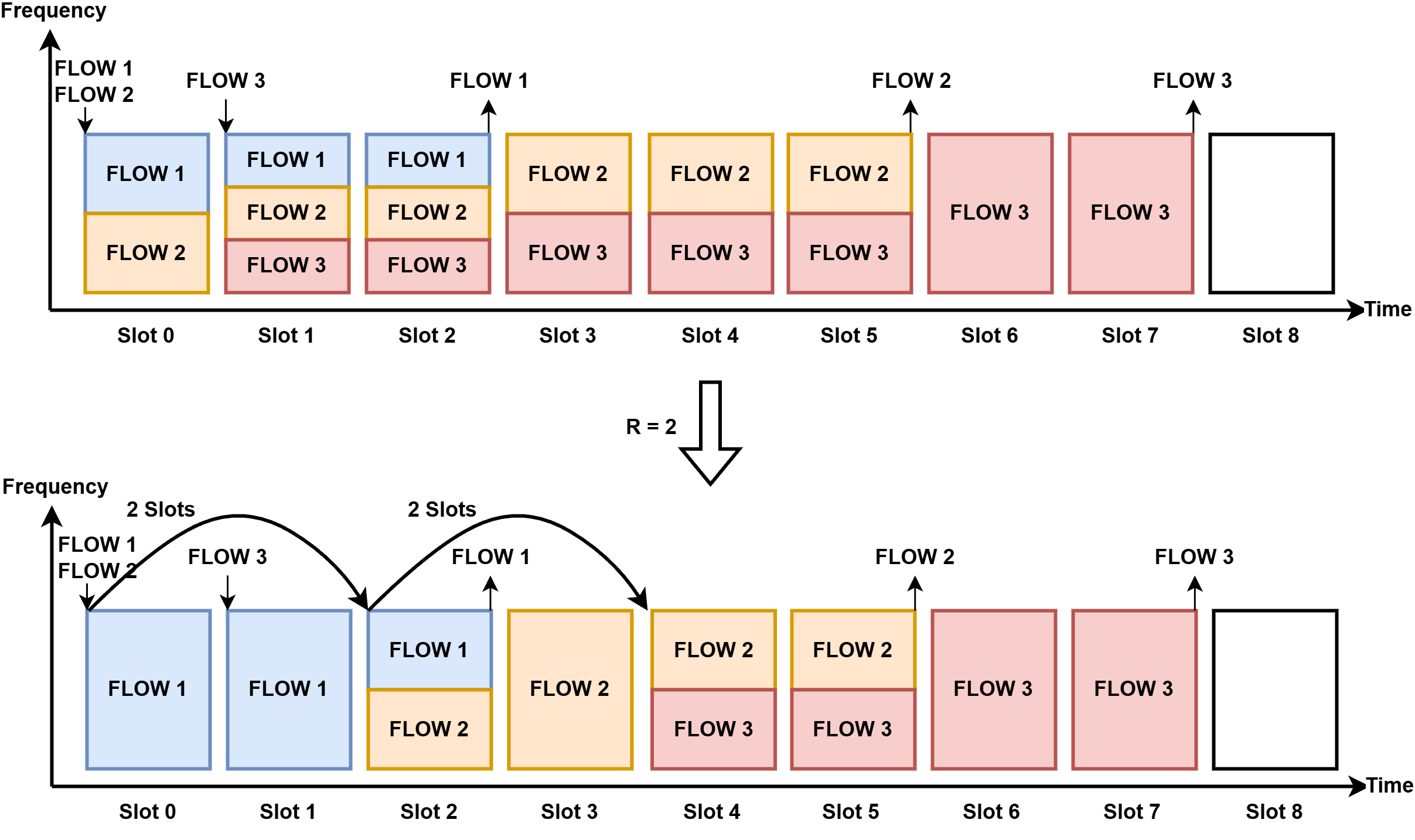} 
\caption{An example of a constructed schedule table} 
\label{fig:schedule_table} 
\end{figure}

Once a feasible offset configuration is obtained, it is transformed into a shared-resource schedule table. As shown in Fig.~\ref{fig:schedule_table}, each slot is configured with a prioritized set of flows. Considering an RTT of \(R\) slots, the interval between packets \(f_{i,j}\) and \(f_{m,n}\) is required to be at least \(\min\{K_{i,j},R\}\). Redundant configured opportunities within this interval are removed to avoid unnecessary resource reservation.

\section{Performance Evaluation}

In this section, we evaluate the schedulability and scalability of the proposed approach through extensive simulations based on industrial use cases specified in 3GPP TS 22.104~\cite{3gpp22104}.

\subsection{Experimental Setup}

\begin{table}[t]
\centering
\caption{Experiment Settings}
\label{tab:exp_setting}
\begin{tabular}{lc}
\hline
Parameter & Value \\
\hline
Slot length & 1 ms \\
Subcarrier spacing & 15 kHz \\
Number of flows & 6--18 \\
Period (ms) & \{4, 5, 8, 10, 12, 15, 20,\\
             & 40, 50, 60, 100, 200, 250\} \\
Reliability requirement & $99.9\% \sim 99.99999\%$ \\
\hline
\end{tabular}
\end{table}

As shown in Table~\ref{tab:exp_setting}, random flow sets are generated based on the performance requirements and configuration parameters of periodic downlink services in 3GPP TS 22.104~\cite{3gpp22104}, following the baseline setup in~\cite{zhang2023}.
For each flow \(f_i\), the period \(T_i\) is randomly selected from Table~\ref{tab:exp_setting}, and the required transmission slots \(K_i\) are determined by the target reliability (``\(x\)-nines''). For example, \(K_i=5\) corresponds to \(99.999\%\) reliability when the per-transmission success probability is 0.9. Consistent with URLLC assumptions for small packets~\cite{liu2022}, each packet occupies one slot, and the RTT is set to 4 slots.

As discussed in Section~\ref{sec:introduction}, this work is, to the best of our knowledge, the first to incorporate proactive HARQ into periodic traffic while providing formal latency and reliability guarantees via schedulability analysis. Existing approaches cannot directly meet these requirements. Therefore, we compare our method with adapted analysis algorithms for commonly used retransmission mechanisms in 5G systems to evaluate schedulability and computational overhead.

\textbf{1) Reactive HARQ:} 
The method in Section~\ref{sec:reactive HARQ} is used to analyze whether all flows satisfy their constraints.

\textbf{2) K-Repetition:}
A method similar to~\cite{zhang2023} is adopted, where SMT is used to solve the problem in Section~\ref{sec:Repetition}.

% \textbf{3) Proactive HARQ (RD-PaS Adaptation):}
% We adapt RD-PaS~\cite{zhang2022reliable} by searching flow offsets and greedily allocating transmission opportunities. Specifically, its multi-hop process is interpreted as consecutive packet transmissions on a shared resource, and additional opportunities are incrementally assigned to each consecutive packet group until the reliability requirements are satisfied or no feasible allocation exists.

\textbf{3) Proactive HARQ:} 
Based on Section~\ref{sec:algorithm}, the proposed two-stage algorithm is used to find feasible configurations.

\subsection{Results for Different Scheduling Mechanisms}

This subsection compares the schedulability ratio and computational overhead of different retransmission mechanisms under their corresponding analysis frameworks. We vary utilization, flow count, HARQ RTT, and success probability to evaluate the robustness and advantages of proposed algorithm.

\subsubsection{Performance under Different Flow-Set Parameters}
\label{subsec:flow_set_parameters}

\begin{figure}[!t]
\centering

\subfloat[Schedulability Ratio]{
    \includegraphics[width=0.48\linewidth]{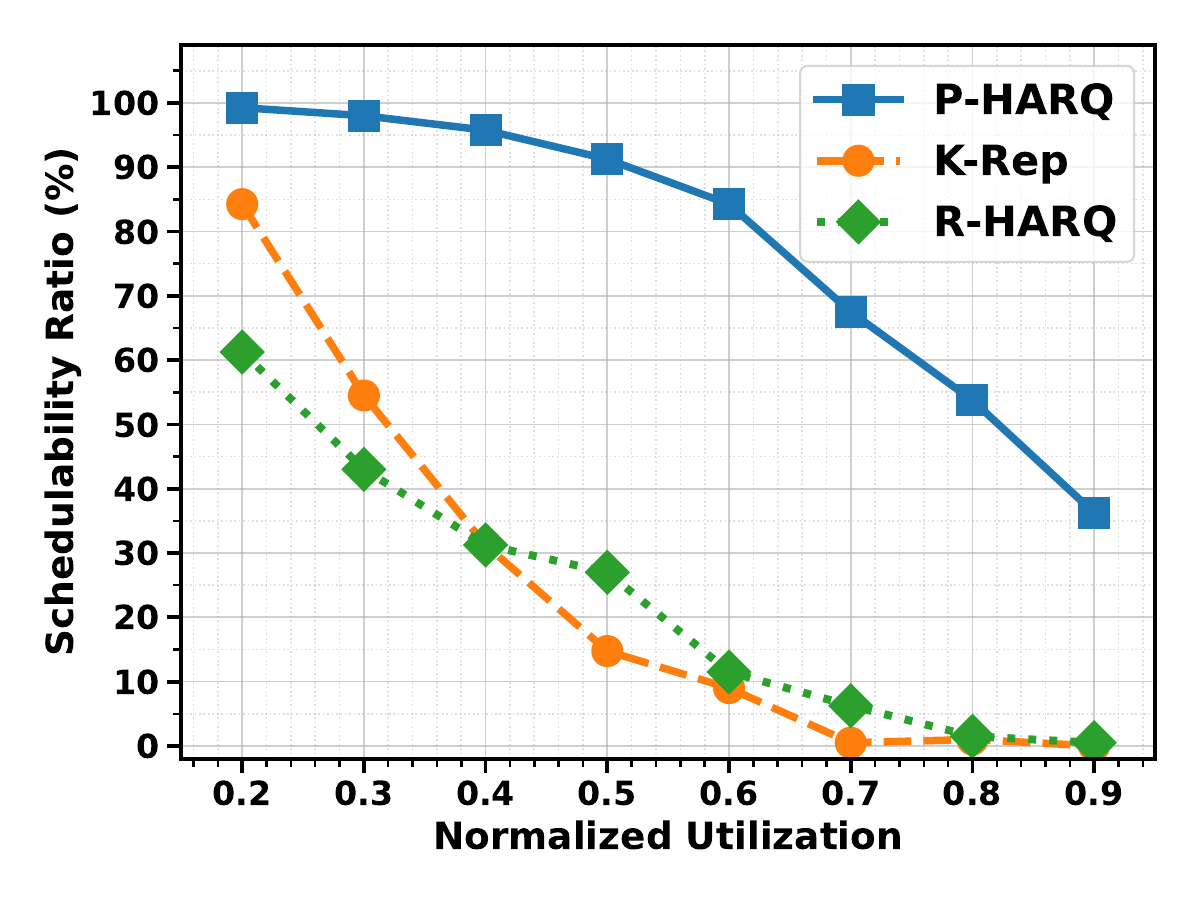}
    \label{fig:schedulability_util}
}
\subfloat[Average Runtime]{
    \includegraphics[width=0.48\linewidth]{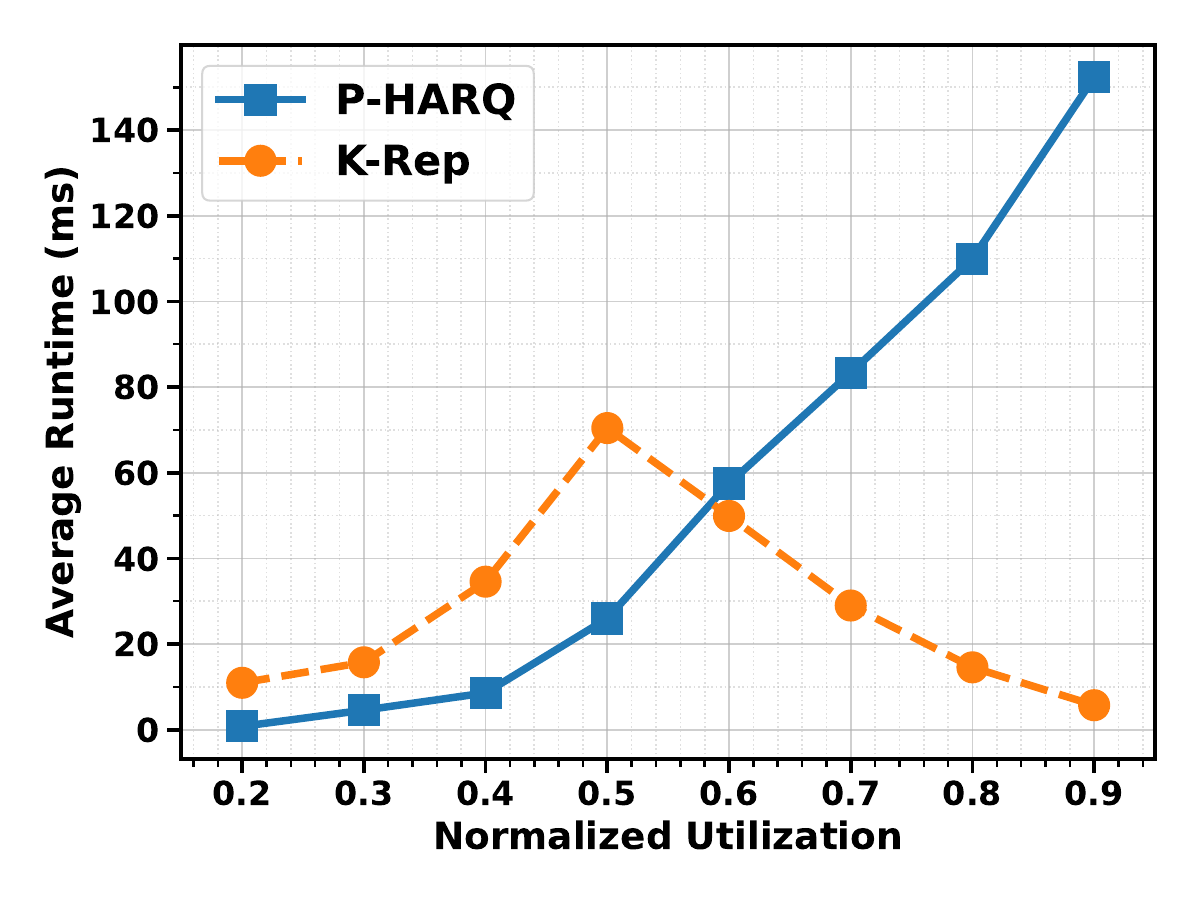}
    \label{fig:runtime_util}
}
\caption{Performance comparison under different utilizations.}
\label{fig:mechanism_comparison}
\vspace{-0.5em}
\end{figure}

\begin{table}[!t]
\centering
\caption{Runtime Comparison under Different Utilization}
\label{tab:runtime_comparison}
\footnotesize
\renewcommand{\arraystretch}{0.80}
\begin{tabular}{c|c|c|c|c}
\hline
Utilization & GA Avg & GA Max & SMT Avg & SMT Max \\
\hline
0.2 & 0.790 & 69.37   & 12.375  & 133.63 \\
0.3 & 4.416 & 400.19  & 25.355  & 370.02 \\
0.4 & 8.597 & 671.14  & 86.696  & 2713.31 \\
0.5 & 24.366 & 1126.55 & 346.877  & 15553.45 \\
0.6 & 57.496 & 1247.50 & 71.140  & 533.10 \\
0.7 & 83.359 & 1559.97 & 70.296  & 440.626 \\
0.8 & 109.932 & 1333.05 & 168.265  & 607.59 \\
0.9 & 152.291 & 4952.43 & ------   & ------ \\
\hline
\end{tabular}
\vspace{-1em}
\end{table}

We first evaluate schedulability and runtime under different flow-set parameters, including utilization and number of flows. The utilization of each flow is defined as \( U_i = \frac{K_i}{T_i} \), where \(K_i\) is the minimum number of required transmission opportunities and \(T_i\) is the period. The total utilization is \(U=\sum_i U_i\). For each setting, 400 flow sets are randomly generated.

Fig.~\ref{fig:schedulability_util} shows the schedulability ratio under different utilization levels. The proposed method consistently achieves the highest schedulability,  and its advantage becomes more pronounced as utilization increases.  At \(U=0.9\), the other methods can hardly schedule any flow set, while the proposed method still achieves 36.25\% schedulability.

Fig.~\ref{fig:runtime_util} reports the average runtime, excluding reactive HARQ since it does not involve iterative solving. At low utilization (\(0.2\)--\(0.5\)), the GA-based method requires about half the runtime of the SMT-based approach. As utilization increases (\(0.6\)--\(0.9\)), the SMT runtime decreases due to early UNSAT termination when most instances are infeasible. In contrast, the GA may spend more time on infeasible instances because more generations are needed to explore the solution space. Table~\ref{tab:runtime_comparison} reports the maximum runtime of the GA-based method and the SMT runtime on satisfiable instances, showing that the proposed method still maintains an advantage in average computational overhead under high utilization.

\begin{figure}[!t]
\centering
\subfloat[Schedulability Ratio]{
    \includegraphics[width=0.48\linewidth]{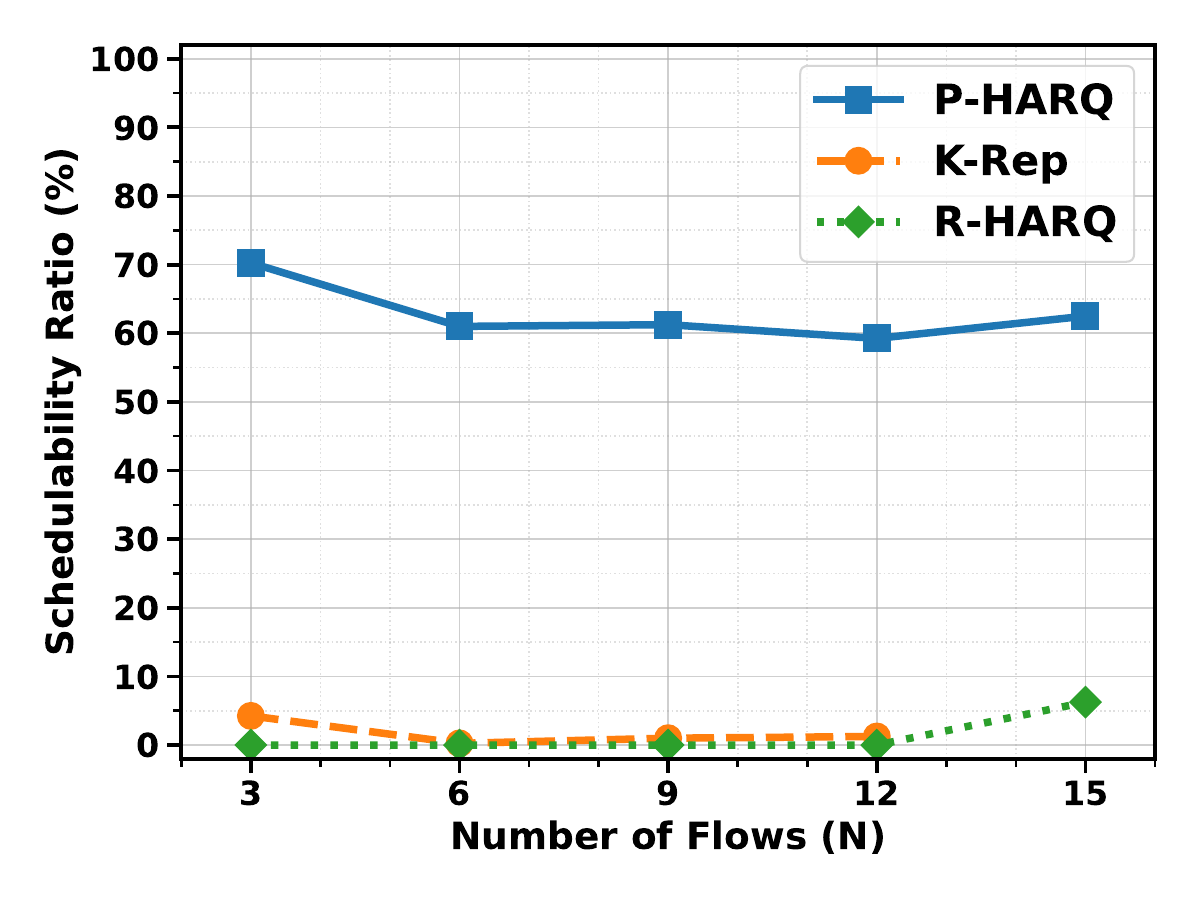}
    \label{fig:schedulability_flow_N_U80}
}
\subfloat[Runtime]{
    \includegraphics[width=0.48\linewidth]{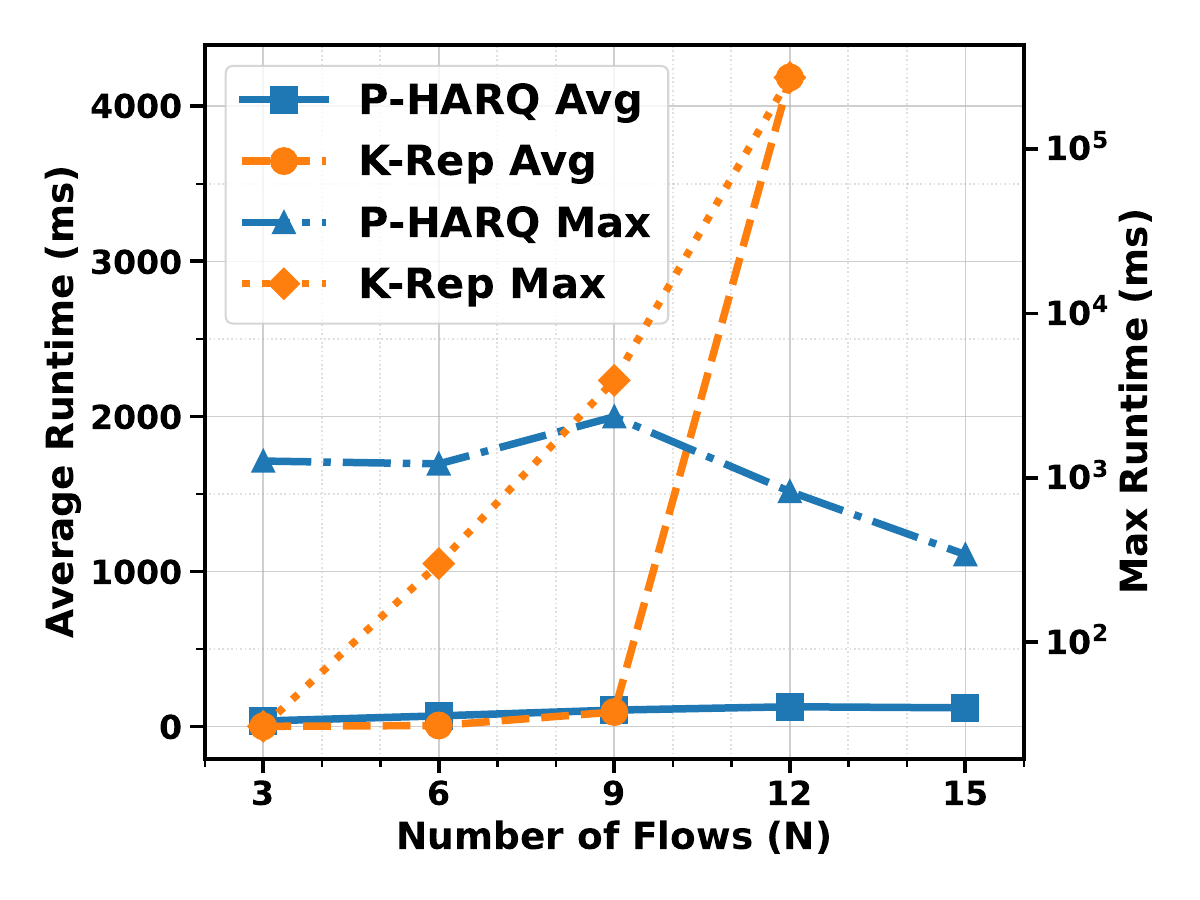}
    \label{fig:runtime_flow_U80}
}
\caption{Performance comparison under different numbers of flows at 80\% utilization.}
\vspace{-1.5em}
\label{fig:flow_comparison_all}
\end{figure}

We further evaluate scalability with respect to the number of flows. The total utilization is fixed at 80\%. As shown in Fig.~\ref{fig:schedulability_flow_N_U80}, both K-Rep and R-HARQ fail to find feasible configurations for most sets, whereas the proposed two-stage algorithm schedules over 60\% of them under proactive HARQ.

Fig.~\ref{fig:runtime_flow_U80} shows the corresponding runtime, including average and maximum values. Since K-Rep relies on SMT solving, its complexity grows rapidly with the number of flows due to increasing contention-free constraints; thus, both average and maximum runtimes increase significantly, with the latter growing by orders of magnitude. When the number of flows reaches 15, the SMT-based method fails to return results within a reasonable time and is omitted. In contrast, the proposed method uses a generation-limited search and is less sensitive to the number of flows: its average runtime remains below 150 ms, and its maximum runtime is bounded within \(10^3\) ms.

\subsubsection{Performance under Different System Parameters}

\begin{figure}[!t]
\centering
\subfloat[GA under different RTTs]{
    \includegraphics[width=0.48\linewidth]{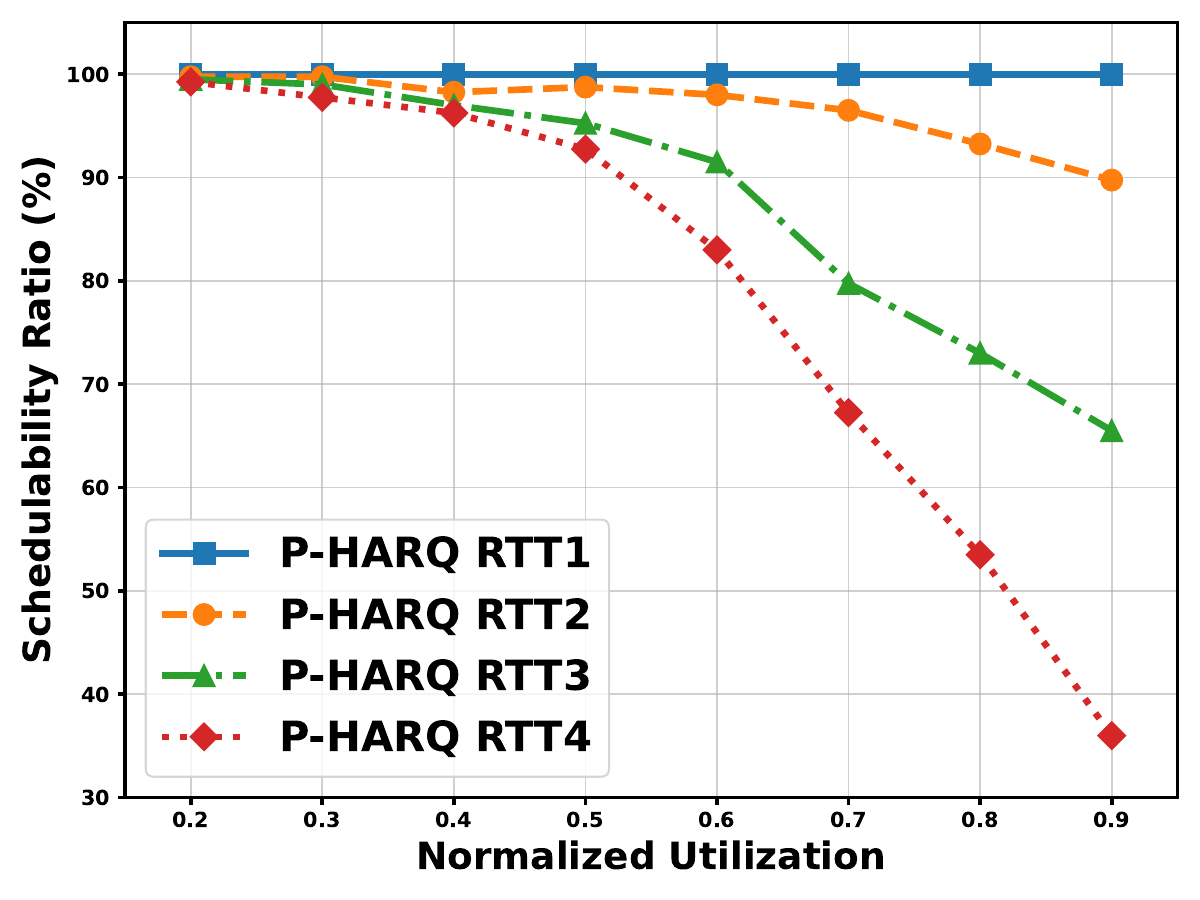}
    \label{fig:ga_rtt}
}
\subfloat[Comparison across mechanisms]{
    \includegraphics[width=0.48\linewidth]{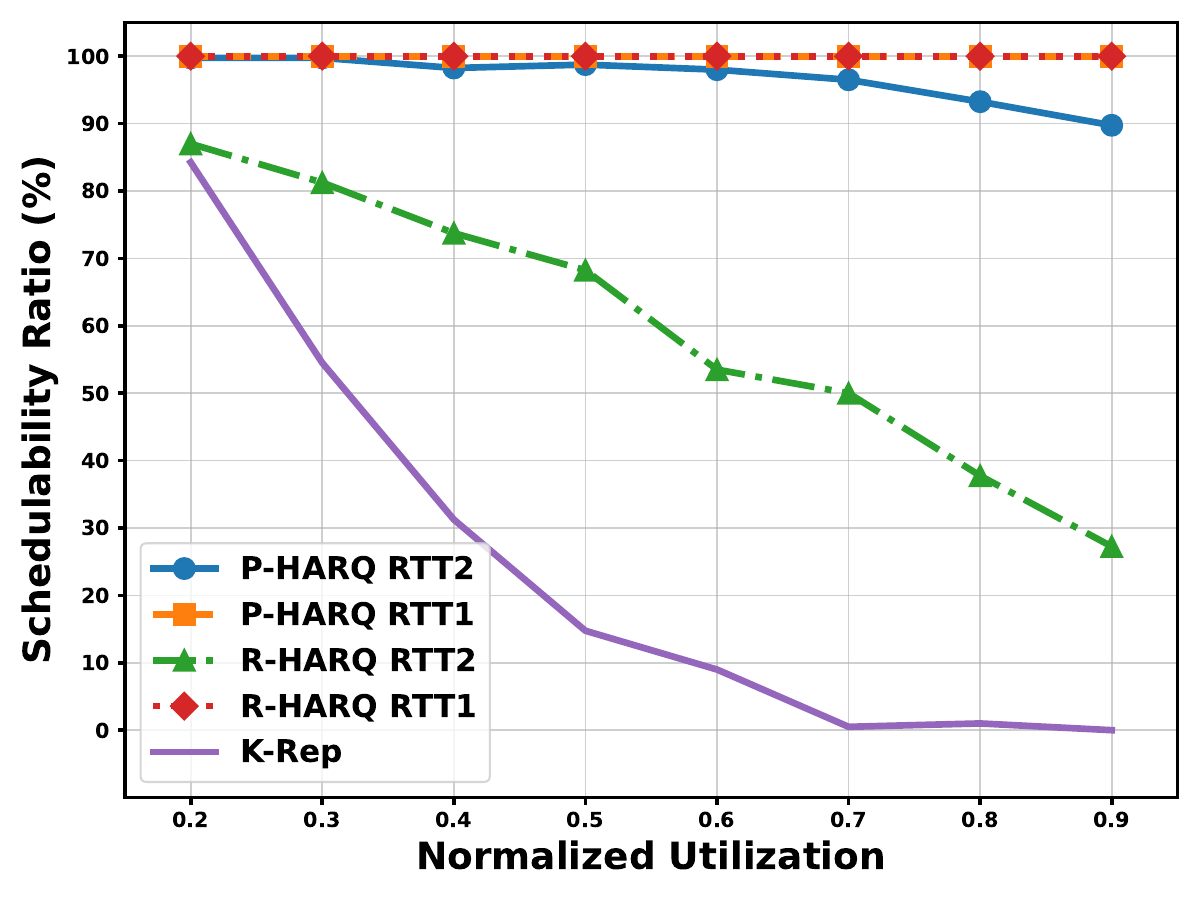}
    \label{fig:rtt_mixed}
}
\caption{Schedulability under different RTT settings.}
\vspace{-1.5em}
\label{fig:rtt_comparison}
\end{figure}

\begin{figure}[!t]
\centering
\subfloat[Worst-case channel with \(p=0.8\)]{
    \includegraphics[width=0.48\linewidth]{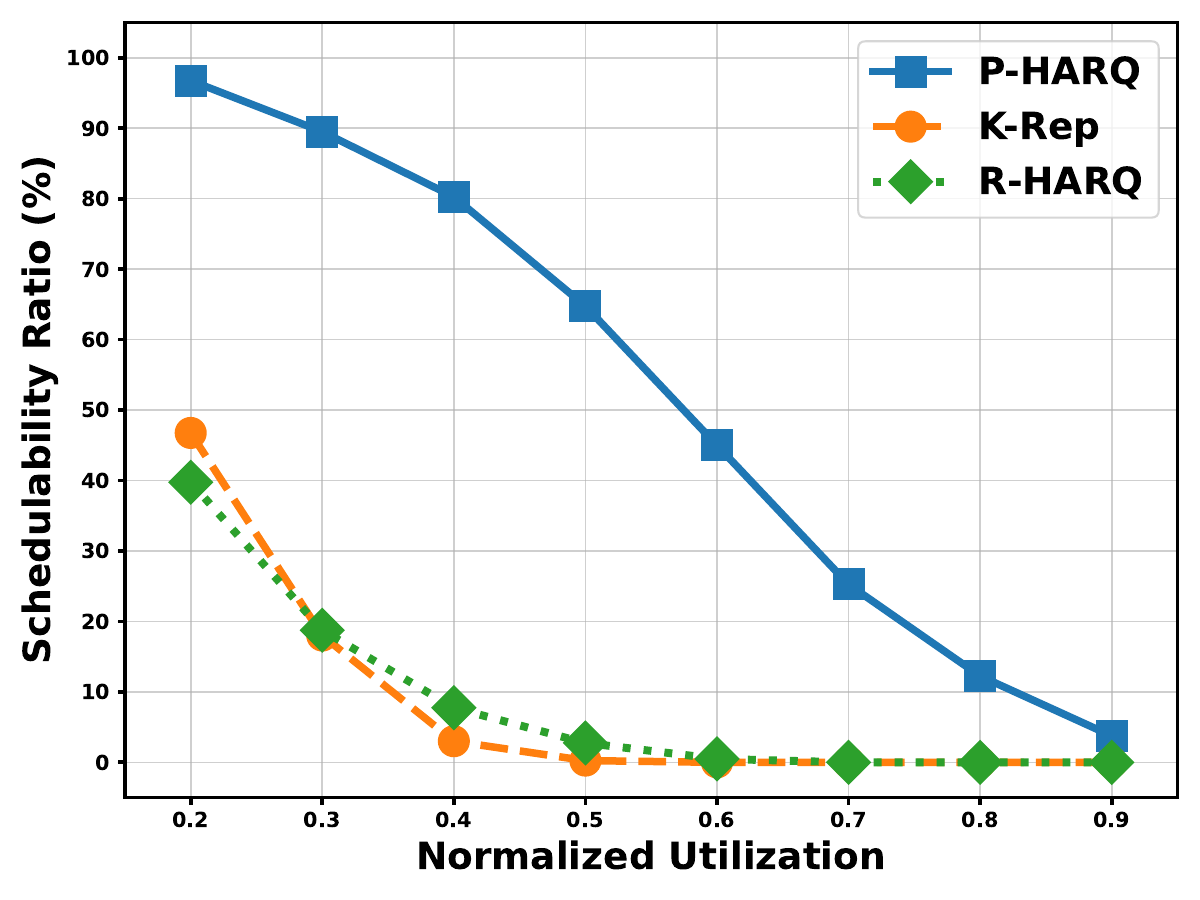}
    \label{fig:p_comparison_fixed}
}
\subfloat[Time-varying fading channel]{
    \includegraphics[width=0.48\linewidth]{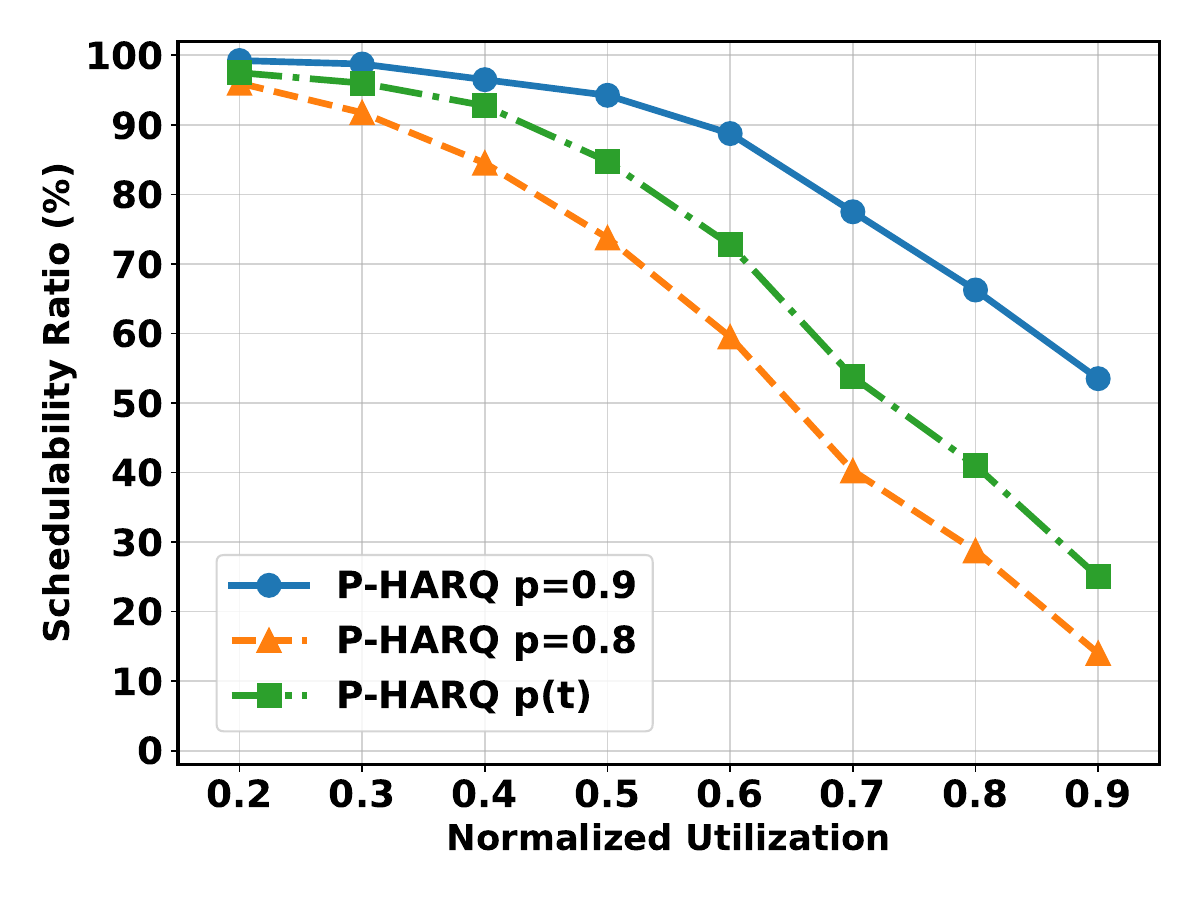}
    \label{fig:p_comparison_fading}
}
\caption{Schedulability under different channel conditions.}
\vspace{-1.5em}
\label{fig:p_comparison}
\end{figure}

In this subsection, we evaluate schedulability under different system parameters, focusing on the impact of HARQ RTT and single-transmission success probability \( p \) across varying utilizations.

We first examine HARQ RTT. While the default RTT is 4 slots, prior work and standards have explored reducing RTT via early feedback~\cite{airod2021harq,strodthoff2019enhanced} and self-contained slot structures in 5G~\cite{qualcomm2018nr}. As shown in Fig.~\ref{fig:ga_rtt}, smaller RTT enables earlier feedback and more efficient reuse of transmission opportunities, improving schedulability; larger RTT delays feedback, causing redundant transmissions and deadline misses.

Fig.~\ref{fig:rtt_mixed} compares different mechanisms. Since K-Rep does not rely on ACK-based retransmissions, it is insensitive to RTT. The proposed method consistently outperforms the baselines, and under the evaluated settings, both mechanisms achieve 100\% schedulability when \(R=1\)..

We next evaluate the impact of the single-transmission success probability \(p\), which is determined by channel conditions. As shown in Fig.~\ref{fig:p_comparison_fixed} and Fig.~\ref{fig:p_comparison_fading}, we compare different algorithms under a worst-case channel with \(p=0.8\) and evaluate P-HARQ under a periodically time-varying fading model caused by CSI aging ~\cite{baddour2005autoregressive}. Harsh channel conditions lead to longer transmission durations, and the schedulability of both baselines drops significantly. In contrast, under both the worst-case channel and the periodic fading channel, the proposed method maintains high schedulability and robustness.

\subsection{Results for the Two-Stage Algorithm}

\begin{figure}[!t]
\centering
\subfloat[Impact of number of generations]{
    \includegraphics[width=0.48\linewidth]{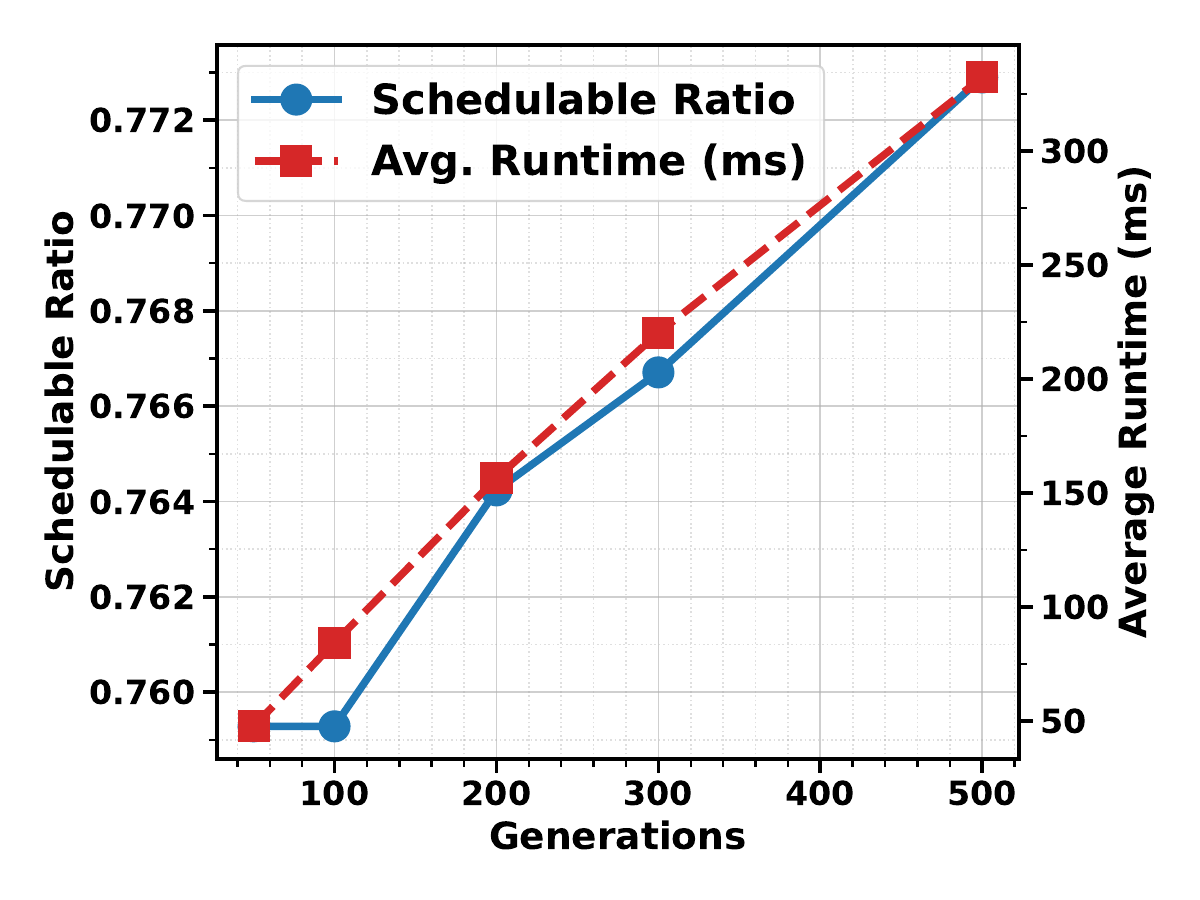}
    \label{fig:ga_gen}
}
\subfloat[Impact of population size]{
    \includegraphics[width=0.48\linewidth]{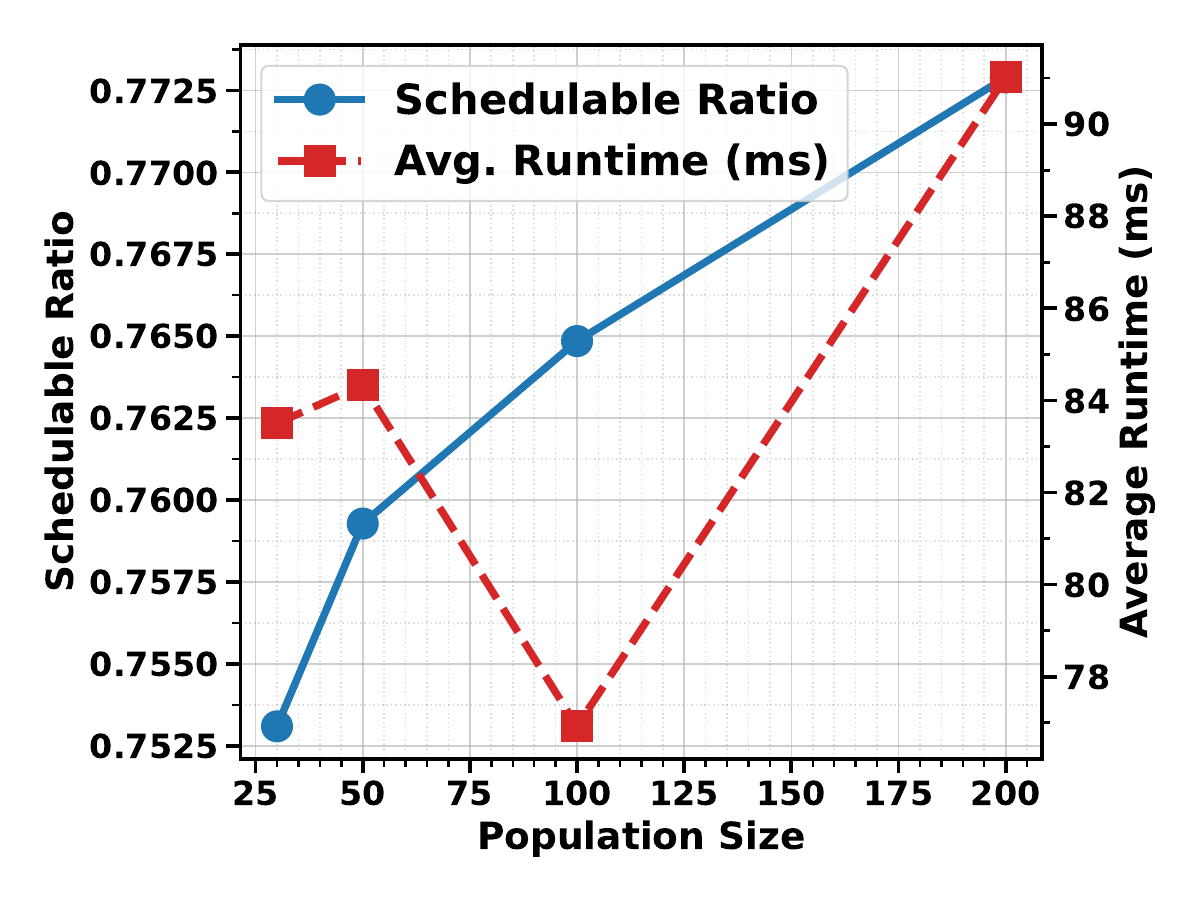}
    \label{fig:ga_pop}
}
\caption{GA parameter impacts on schedulability and runtime.}
\vspace{-1.5em}
\label{fig:ga_param}
\end{figure}

\begin{figure}[!t]
\centering
\subfloat[Schedulability contribution]{
    \includegraphics[width=0.48\linewidth]{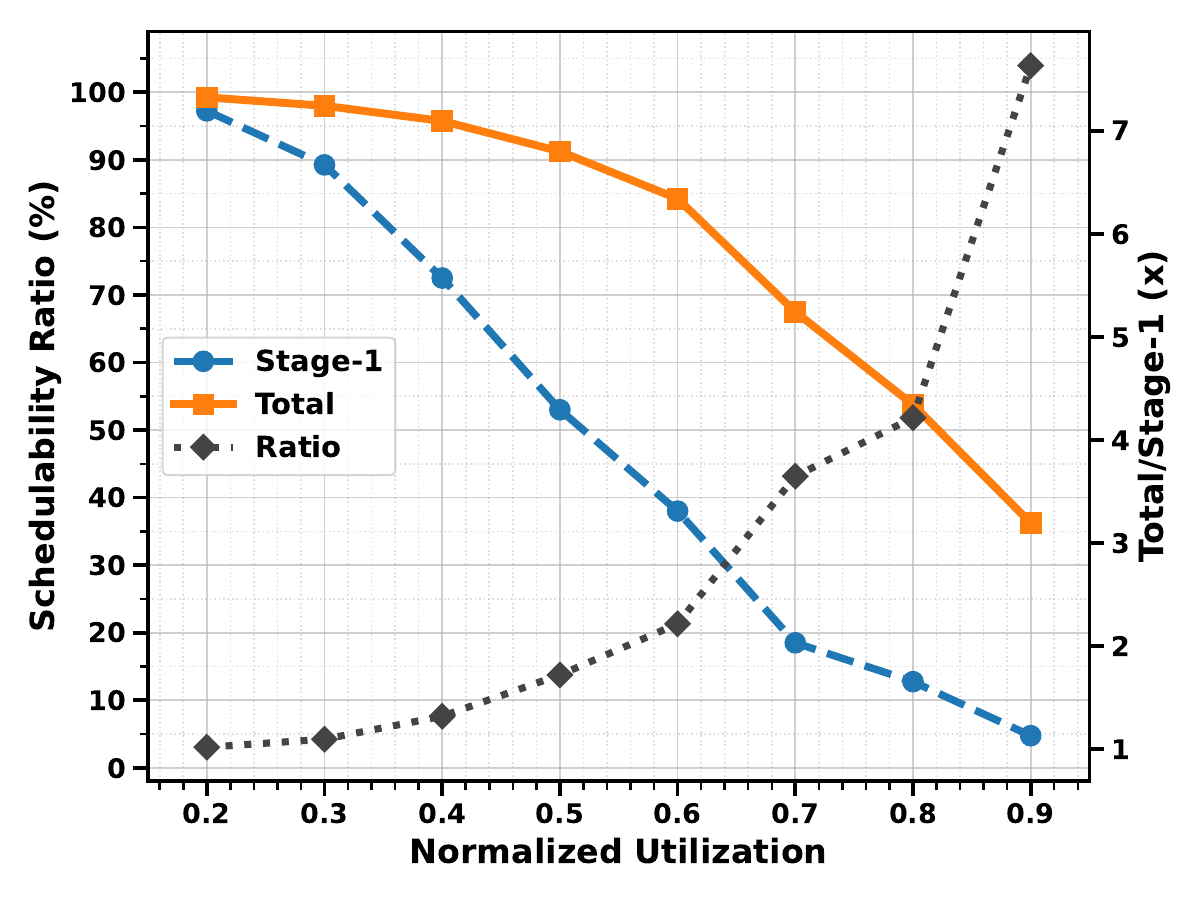}
    \label{fig:two_stage_sched}
}
\subfloat[Runtime comparison]{
    \includegraphics[width=0.48\linewidth]{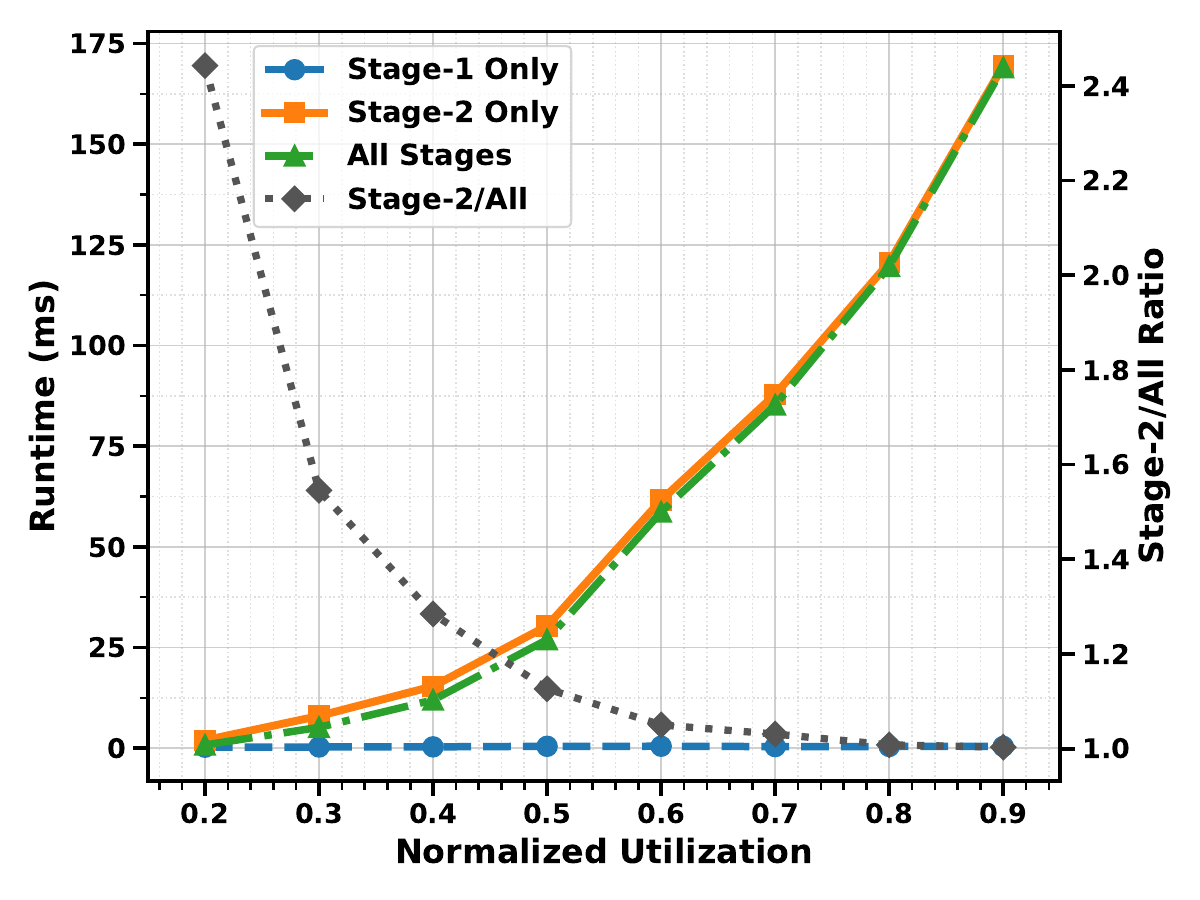}
    \label{fig:two_stage_runtime}
}
\caption{Schedulability and runtime of two-stage algorithm.}
\vspace{-1.5em}
\label{fig:two_stage_comparison}
\end{figure}

In this subsection, we evaluate the proposed two-stage algorithm from two aspects: the impact of GA parameters and the contribution of each stage.

\subsubsection{Performance under Different Algorithm Parameters}

We first evaluate the impact of key GA parameters, namely the number of generations and population size, on schedulability and runtime. Experiments are conducted at 80\% utilization, where differences are more pronounced.

As shown in Fig.~\ref{fig:ga_gen}, increasing the number of generations generally improves schedulability at the cost of higher runtime. However, the improvement is not monotonic; for instance, increasing from 50 to 100 generations yields little gain, as both settings explore a limited portion of configuration space.

Fig.~\ref{fig:ga_pop} shows that larger population sizes improve schedulability by increasing solution diversity. Notably, a population size of 100 results in lower runtime than 50, indicating that small populations may lead to inefficient search and slower convergence due to limited exploration and local optima.

Overall, larger generations and populations improve schedulability at the cost of computational overhead, highlighting the need for proper parameter tuning. Nevertheless, even with relatively small parameter values, the proposed method maintains a clear schedulability advantage over baseline approaches.

\subsubsection{Contribution of the Two Stages}

% Fig.~\ref{fig:two_stage_sched} compares stage-1, which uses worst-case analysis only, with the full two-stage algorithm, where stage-2 applies GA-based offset search after stage-1 fails. Stage-1 alone schedules most flow sets at low utilization, while stage-2 becomes essential as utilization increases; together, the two stages improve schedulability by over seven times compared with stage-1 alone. 
% Fig.~\ref{fig:two_stage_runtime} shows that stage-1 achieves up to a \(2.44\times\) speedup at low utilization, while its overhead becomes negligible at high utilization. Thus, the two stages are complementary: stage-1 quickly handles easy instances with low overhead, while stage-2 schedules harder flow sets when stage-1 fails.

Fig.~\ref{fig:two_stage_sched} shows the fraction of flow sets directly handled by stage~1 and the additional schedulability gained from stage~2. Stage~1 succeeds mainly at low utilization, while GA-based offset search becomes increasingly important as utilization grows. Fig.~\ref{fig:two_stage_runtime} shows that stage~1 reduces runtime by up to \(2.44\times\) for easy instances, with negligible overhead otherwise. Thus, the two stages are complementary.

\subsection{Simulation-Based Case Study}

\begin{figure}[!t]
\centering
\includegraphics[width=\linewidth]{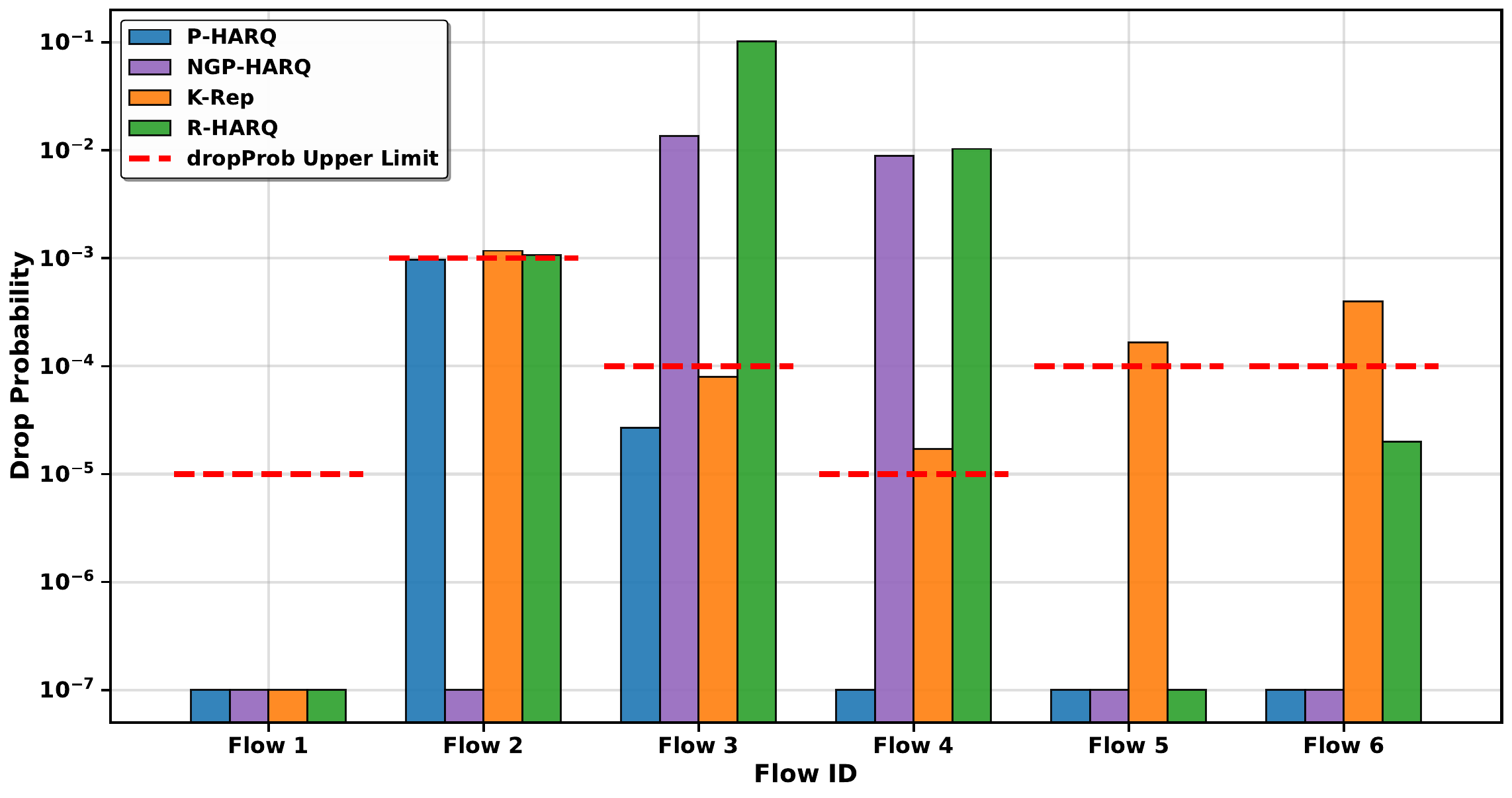}
\caption{Packet drop probability comparison across different retransmission mechanisms.}
\vspace{-0.5em}
\label{fig:case_study}
\end{figure}

\begin{table}[t]
\centering
\caption{Reliability Performance Comparison of Different Mechanisms}
\label{tab:case_study_reliability}
\begin{tabular}{c|c|c|c|c|c}
\hline
\textbf{Flow} & \textbf{Target} & \textbf{P-HARQ} & \textbf{NGP-HARQ} & \textbf{K-Rep} & \textbf{R-HARQ} \\
\hline
Flow 1 & 0.999990 & 1.000000 & 1.000000 & 1.000000 & 1.000000 \\
Flow 2 & 0.999000 & 0.999027 & 1.000000 & 0.998827 & 0.998933 \\
Flow 3 & 0.999900 & 0.999973 & 0.986453 & 0.999920 & 0.898187 \\
Flow 4 & 0.999990 & 1.000000 & 0.991083 & 0.999983 & 0.989683 \\
Flow 5 & 0.999900 & 1.000000 & 1.000000 & 0.999833 & 1.000000 \\
Flow 6 & 0.999900 & 1.000000 & 1.000000 & 0.999600 & 0.999980 \\
\hline
\end{tabular}
\vspace{-1.5em}
\end{table}

In this subsection, we conduct a simulation-based case study using the MATLAB 5G Toolbox~\cite{mathworks5g}. We implement three retransmission mechanisms and their corresponding analytical scheduling frameworks. We also include non-guaranteed proactive HARQ (NGP-HARQ), which uses the P-HARQ transmission mechanism with a configuration that has not passed the proposed schedulability test.

To enhance realism, we construct a flow set based on Mobile Operation Panel profiles in 3GPP TS 22.104~\cite{3gpp22104}, including use cases such as remote control and emergency stop. The flow set contains six periodic flows. With a slot duration of 0.5 ms, the system utilization reaches 93.7\%.
Under this high-utilization setting, no feasible configurations can be found for R-HARQ or K-Rep. Therefore, best-effort configurations are used: R-HARQ defers transmissions under contention, while K-Rep drops lower-priority transmissions. For P-HARQ, we use the proposed two-stage algorithm and construct the schedule table following Section~\ref{sec:alg2}. NGP-HARQ uses an unverified configuration and follows the same P-HARQ transmission procedure.

We simulate continuous transmissions over 10 minutes (1{,}200{,}000 slots) with a per-transmission success probability of 0.9. A packet is considered dropped if all allocated transmission opportunities are exhausted without successful decoding.

Fig.~\ref{fig:case_study} and Table~\ref{tab:case_study_reliability} show that P-HARQ scheduled by our algorithm does not achieve the lowest drop probability for every flow, but it is the only case satisfying all reliability requirements. In contrast, R-HARQ, K-Rep, and NGP-HARQ each violate the target of at least one flow, demonstrating the importance of the proposed analysis and configuration: the high schedulability is not provided by the P-HARQ mechanism alone, but by its formal guarantees under our framework.

The proposed method also improves resource efficiency. Under K-Rep, the total number of transmission attempts is 1{,}124{,}500, occupying 93.71\% of slots. P-HARQ reduces this to 779{,}400, occupying 64.95\% of slots and saving about 30\% of transmissions. This reduction leaves more resources for aperiodic or stochastic traffic and can improve their reliability and latency during random access.

Overall, through formal analysis and schedule construction, the proposed method provides guaranteed scheduling for heterogeneous periodic flows within acceptable time.

\section{Conclusion}

This paper presents a DTMC-based framework for analyzing heterogeneous periodic flows under proactive HARQ, capturing both reliability and latency.We further develop a two-stage scheduling algorithm that combines synchronous-release testing with genetic offset search to improve schedulability and computational efficiency. Evaluation results show clear advantages over existing mechanisms and algorithms. Future work will extend the framework to OFDMA-based multi-slice systems with dynamic resource allocation.

\bibliographystyle{IEEEtran}
\bibliography{refs}

\end{document}